\documentclass[reprint, aps,prd,twocolumn,showpacs,showkeys, 10pt, longbibliography, nolinenumbers, floatfix]{revtex4-2}

\usepackage{amsmath}
\usepackage{graphicx}
\usepackage{float}
\usepackage{dcolumn}
\usepackage{multirow}
\usepackage{natbib}
\usepackage{changes}
\usepackage[colorlinks = true,linkcolor = blue,urlcolor  = blue,citecolor = blue,anchorcolor = blue]{hyperref}
\usepackage{enumerate}
\usepackage{verbatim}
\usepackage{orcidlink}

\begin{document}

\title{ Machine and Deep Learning Regression for Compact Object Equations of State}

\author{I. Stergakis}

\author{Th. Diakonidis}

\author{Ch.C. Moustakidis}


\affiliation{Department of Theoretical Physics, Aristotle University of Thessaloniki, 54124 Thessaloniki, Greece  }

\begin{abstract}
One of the outstanding challenges in nuclear physics is achieving a comprehensive understanding of the equation of state (EoS) that governs dense nuclear matter, and, by extension, the internal composition and structure of compact astrophysical objects such as neutron stars and quark stars. The primary approach to addressing this problem has thus far been predominantly theoretical, involving the construction of equations of state based on plausible nuclear and particle physics theories. These theoretical predictions are subsequently evaluated against empirical results, which are obtained primarily from heavy-ion collision experiments and, increasingly, from astrophysical observations of compact objects. In recent years, several alternative analytical methodologies have been developed that integrate the aforementioned theoretical and empirical approaches. These methods incorporate advanced data analysis and processing techniques, including Bayesian inference, and machine learning and deep learning frameworks, to enhance the interpretation and constraining of the equation of state for dense nuclear matter. In this study, the aforementioned techniques, particularly machine learning and deep learning, are employed with the objective of enhancing our understanding of the equation of state by leveraging information derived from mass-radius diagrams of compact objects. More specifically, this study utilizes a large dataset of equations of state constructed using multimodal formulations of neutron stars, along with a corresponding dataset for quark stars. These equations of state are consistent with known physical constraints and have been developed to reflect the highest degree of physical plausibility. Through the application of machine learning and deep learning techniques, it becomes possible to infer or reconstruct the underlying equations of state based on input from mass-radius diagram data.


\end{abstract}

\maketitle

\section{Introduction}
Neutron stars are among the most important natural laboratories beyond Earth for the study of physics~\cite{Shapiro:1983du,Haensel2007NeutronS1,Glendenning-2000,schaffner-bielich_2020}. The extreme conditions they harbor including  density, temperature, magnetic fields, gravity, and more  cannot, at least for now, be recreated in terrestrial laboratories. Yet, they  provided us with a wealth of insights into the laws of nature, which must still be analyzed and explained using the physics we already understand.
Nevertheless, the limited knowledge of the composition and behavior of dense nuclear matter, both experimentally and theoretically, makes the study of compact stars very challenging. On the one hand, as noted previously, experimental data are unavailable or very limited, at best. On the other hand, theoretical predictions are subject to considerable uncertainty as current models are reliable only within the regime of relatively low densities.

In recent years, however, the wealth of observations and related studies of neutron stars, combined with the groundbreaking detection of gravitational waves from the mergers of binary compact systems, has reignited scientific interest in the field. Most importantly, these developments have given us the opportunity to reexamine and refine existing theoretical models.
Beyond the interplay of theory and observation, there is also a third avenue for advancing our understanding of neutron stars: one that integrates the two. Statistical analysis has in recent years emerged as a particularly powerful approach, with methods such as Bayesian analysis and machine learning being applied with considerable success~\cite{Read-2009,raithel2016neutron,Ozel-2009,Jiang-2023,Zhou-2023,Huth-2022,Peter-2024,Patra-2025,Imam-2024,Patra-2022,Imam-2022,Clement-2025,Fujimoto-2018,Fujimoto-2020,Thakur-2024,Krastev-2023,Guo-2024,Chatterjee-2024,Zhou-2023b,Soma-2024,Li-2025,Bejger-2025,Papigkiotis-2025,Ventagli-2025,Fujimoto-2024,Carvalho-2024a,Carvalho-2024b,Brandes-2024,Fujimoto-2021,Ferreira-2021}.

The primary objective of the present study is to develop a computational framework capable of reproducing the state-of-the-art understanding of dense nuclear matter in neutron stars (or quark stars), based on the known mass–radius relationship. In particular, the relativistic inverse stellar structure problem determines the equation of state of the stellar matter
given a knowledge of suitable macroscopic observables (e.g. their masses and radii) of the stars composed of that material~\cite{Lindblom-1992,Lindblom-2025}.
This approach in fact is the inverse of the conventional methodology, where the equation of state is first specified and employed to predict the mass–radius relation of the star. It should be noted that previous attempts have been made to address this problem through various approaches (see Lindblom et al.~\cite{Lindblom-1992,Lindblom-2025,Lindblom-2010,Lindblom-2018,Lindblom-2022}). In the present study, we employ an alternative approach to reconstruct the equation of state from the mass–radius relation utilizing machine learning and deep learning techniques.

In particular,  we employ the method discussed in reference~\cite{Read-2009} to construct a comprehensive dataset of equations of state for the neutron star core, utilizing polytropic  equations of state. These equations of state span a broad region in the pressure–energy density diagram, thereby encompassing as much information as possible about neutron star matter across both high and low densities. In the second step, we use this dataset to construct the corresponding M–R diagrams, which span the widest possible range of masses and radii. 
Using machine learning and deep learning methods (regression), we train our system to reconstruct the underlying equation of state from a given mass–radius relation, effectively reproducing it through the inverse process. The ultimate goal of this work is to reconstruct the corresponding equations of state from the mass–radius relations inferred through potential astrophysical observations.

Regression by definition is a statistical method for quantifying the relationship between independent input variables (usually called features or predictors) and dependent output variables (responses or targets). 
In our case, the regression task is formulated as follows. The input consists of 8 or 16 mass–radius pairs spanning the M–R space of neutron stars, which serve as predictors for the inverse supervised problem mentioned above. The outputs are twelve energy cases. We term this problem “inverse” because the forward (direct) problem is usually solved by integrating the Tolman–Oppenheimer–Volkoff (TOV) equations, which connect the equation of state to observable M–R relations~\cite{Tolman1939, Oppenheimer1939}.
We address the inverse regression problem using two approaches: (i) conventional machine learning regression models, and (ii) a deep learning artificial neural network.
An analogous methodology is applied to the case of quark stars. Specifically, we generate a comprehensive dataset of equations of states based on the Color–Flavor–Locked (CFL) model, from which we calculate the corresponding mass–radius diagrams. This dataset is subsequently employed to train our model to operate (predict) in the inverse direction.

The reconstruction task is addressed using a set of machine learning algorithms designed for multivariate multiple regression. In particular, we employ \textsc{Decision Trees} \cite{CART},
\textsc{Random Forests} \cite{Breiman2001}, \textsc{Gradient Boosting} \cite{Friedman2001, Natekin2013}, \textsc{XGBoost} \cite{Chen2016}, and \textsc{Deep Learning} \cite{Hornik1989}. These methods span a spectrum from simple and interpretable baselines to highly expressive nonlinear models. \textsc{Decision Trees} provide a transparent starting point, while ensemble methods such as \textsc{Random Forests} and \textsc{Gradient Boosting} reduce variance and bias to achieve higher accuracy. \textsc{XGBoost} extends boosting with regularization and computational efficiency, and \textsc{Deep Learning} through artificial neural networks enables the learning of complex nonlinear dependencies when sufficient data are available. Their established Python implementations, predictive accuracy, and complementary strengths make them particularly well suited to our problem, in order to gradually succeed good accuracy, stability and interpretability to our models.

The paper is organized as follows: Section II is dedicated to the construction of the equation of state of neutron stars and quark stars. Section III presents the statistical methods applied, while in Section IV  the results are presented and discussed. Finally, Section V contains the scientific remarks of the study.

\section{Equation of state}\label{EOS_theory}
\subsection{Neutron stars }
\label{PolyLin_EOSs}
In the present analysis, apart from the diversity of the equations of state, an extensive coverage of the M–R plane is a crucial aspect, as it ensures the generation of a substantial dataset and enhances the generality of the regression models to be developed. To fulfill these objectives, a systematic approach must be adopted involving the artificial generation of a sufficiently large number of distinct equations of state. For this purpose, we employ polytropic equations of state. In particular, we use the method established  in Ref.~\cite{Read-2009} (and applied also in Refs.~\cite{raithel2016neutron,Ozel-2009}). In this approach, the region of mass density, between two boundaries $\rho_{min}$ and $\rho_{max}$, must be chosen and divided into $n$ segments. The EoS is then parametrized in terms of $n$ piecewise polytropes. Setting the values of mass density and pressure at the bounds of polytropic segments' as $\rho_i$ and $P_i$, respectively, each segment is given by \cite{Read-2009}
\begin{equation}\label{press_poly}
\begin{matrix}
    P=K_i\rho^{\Gamma_i} & (\rho_{i-1}\leq\rho\leq\rho_i)
\end{matrix}
\end{equation}
where the value of the constant $K_i$, is determined from the pressure and mass density at the previous fiducial point as follows
\begin{equation}\label{K-poly}
    K_i=\frac{P_{i-1}}{\rho^{\Gamma_i}_{i-1}}=\frac{P_i}{\rho^{\Gamma_i}_i}
\end{equation}
and the polytropic index of the segment $\Gamma_i$, is given by
\begin{equation}\label{Gamma-poly}
    \Gamma_i=\frac{\log_{10}(P_i/P_{i-1})}{\log_{10}(\rho_i/\rho_{i-1})}
\end{equation}
The value of $\Gamma_i$ on each segment is usually chosen arbitrarily. The formulas of the polytropic EoSs are then obtained by integrating the equation
\begin{equation}
d\left(\frac{\epsilon}{\rho}  \right)=-Pd\left(\frac{1}{\rho}  \right)   
\label{dE-p} 
\end{equation}
which for the case $\Gamma\neq1$ leads to 
\begin{equation}
\label{polyEoSGn1form}
    \epsilon(\rho)=(1+a)\rho c^2+\frac{K}{\Gamma-1}\rho^{\Gamma}
\end{equation}
where $a$ is an integration constant. The value of $a$ is determined by requiring the continuity of the EoS along any mass density section at either endpoint and Eq.~(\ref{polyEoSGn1form}) becomes
\begin{eqnarray}\label{poly_EoS_Gn1_gen_form}
&&\epsilon(\rho)=\left[\frac{\epsilon(\rho_{i-1})}{\rho_{i-1}}-\frac{P_{i-1}}{\rho_{i-1}(\Gamma_i -1)}\right]\rho + \frac{K_i}{\Gamma_i-1}\rho^{\Gamma_i}, \nonumber \\
&& \text{ }\text{ } (\rho_{i-1}\leq\rho\leq\rho_i)
\end{eqnarray}
where $K_i$ and $\Gamma_i$ are calculated  
from Eqs.~(\ref{K-poly}) and ~(\ref{Gamma-poly}), respectively.

In the same way, integrating Eq.~(\ref{dE-p}), for the case  $\Gamma=1$, the energy density reads \cite{raithel2016neutron}
\begin{eqnarray}\label{poly_EoS_G1_gen_form}
&&\epsilon(\rho)=\frac{\epsilon(\rho_{i-1})}{\rho_{i-1}}\rho + K_i\ln\left(\frac{1}{\rho_{i-1}}\right)\rho-K_i\left(\frac{1}{\rho}\right)\rho, \nonumber\\
&&\text{ }\text{ } (\rho_{i-1}\leq\rho\leq\rho_i)
\end{eqnarray}

We can rewrite 
Eqs.~(\ref{poly_EoS_Gn1_gen_form}) and ~(\ref{poly_EoS_G1_gen_form}) with the energy density being a function of pressure $\epsilon(P)$. To do so, we use the polytropic relation between pressure and mass density from Eq.~(\ref{press_poly}). For $\Gamma\neq1$, we have
\begin{eqnarray}\label{poly_EoS_Gn1_gen_form2}
&&\epsilon(P)=\left[\frac{\epsilon(\rho_{i-1})}{\rho_{i-1}}-\frac{P_{i-1}}{\rho_{i-1}(\Gamma_i -1)}\right]\left(\frac{P}{K_i}\right)^{\Gamma_i^{-1}}+\frac{P}{\Gamma_i-1}, \nonumber\\
&&\text{ }\text{ } (P_{i-1}\leq P\leq P_i)
\end{eqnarray}
and for $\Gamma=1$
\begin{eqnarray}\label{poly_EoS_G1_gen_form2}
&&\epsilon(P)=\frac{\epsilon(\rho_{i-1})}{\rho_{i-1}}\frac{P}{K_i}+\ln\left(\frac{1}{\rho_{i-1}}\right)P-\ln\left(\frac{K_i}{P}\right)P, \nonumber \\
&&\text{ }\text{ } (P_{i-1}\leq P\leq P_i)
\end{eqnarray}
It is worth noting here that, in the present work, the equations of state have been constructed in such a way as to ensure that causality is not violated. To be more specific, the fundamental requirement of causality derived from special relativity dictates that the pressure gradient of any EoS has an upper bound of 
\begin{equation}\label{caus_limit}
    \frac{dP}{d\epsilon}\equiv\left(\frac{c_s}{c}\right)^2\leq1
\end{equation}
known as the causality limit, where $c_s/c$ is the speed of sound $c_s$ in units of light speed $c$. Thus, according to the above, the equations of state have been constructed in such a way as to ensure that this principle is not violated under any circumstances. This confers greater physical credibility to the technically generated equations of state and, more importantly, it guarantees the correct behavior at high densities.

It is quite possible though, that some of the mock polytropic EoSs, we create, violate the condition of Eq.~(\ref{caus_limit}) after a certain value of mass density $\rho_{tr}$ (or equivalently pressure $P_{tr}$). In those cases, we assume the transition from the polytropic parametrized EoS to an EoS of linear behavior, at pressure $P_{tr}$ and beyond. A Maxwell construction is well-suited to describe this kind of transitions \cite{laskos2025speed}
\begin{equation}\label{Maxwell_construct}
    \epsilon(P) = \begin{cases}
    \epsilon_{Hadron}(P),&P\leq P_{tr} \\
    \epsilon(P_{tr}) +\Delta\epsilon+(c_s/c)^{-2}(P-P_{tr}), &P>P_{tr}
    \end{cases}
\end{equation}
where $\epsilon_{Hadronic}(P)$ in the first line of Eq.~(\ref{Maxwell_construct}) stands for the hadronic phase before the transition, governed by a continuous EoS (polytropic or other), and the second line refers to the maximally stiff high density phase. Note that there is no mixed phase region (as in Gibbs construction) and that the two phases co-exist only at the phase transition pressure $P_{tr}$. Lastly, the term $\Delta\epsilon$ establishes the discontinuity of the energy density and hence the discontinuity of the total EoS at the transition density  $\rho_{tr}$ (or pressure $P_{tr}$). However, in our case we consider a continuous total EoS, so the term $\Delta \epsilon$ vanishes. Furthermore, we impose the extreme causality condition given by Eq.~(\ref{caus_limit}), namely, we set the slope of the linear equation of state to satisfy
$(c_s/c)^{-2}=1$.

Finally, in the present study we employed as a test set a number of equations of state that were derived using either microscopic or phenomenological models, and which have been extensively applied in the literature to the study of neutron stars (see Ref.~\cite{Koliogiannis-2020} and references therein). More specifically, the target was to assess the accuracy of the proposed method in reproducing these equations of state, under the assumption that the corresponding M–R relation they predict is known. We ensured that the selection of these equations of state spans a wide range of the M–R diagram so as to encompass as much as possible the diversity of existing predictions. A common feature among them is the prediction of a maximum mass that exceeds two solar masses.

\subsection{Quark stars: The MIT Bag Model}
The simplest phenomenological model to describe the matter of quark stars is the well-known bag model~\cite{Haensel2007NeutronS1,schaffner-bielich_2020}. 
The Massachusetts Institute of Technology (MIT) bag model considers a gas of free relativistic quarks that are confined within a bag stabilized
by some pressure $B$ from the outside. The quantity B is assumed to be constant and usually dubbed the MIT bag constant~\cite{schaffner-bielich_2020}. In this case, the equation of state is a linear relationship between pressure and energy density and has the form
\begin{equation}\label{MIT_EOS_press}
P=\frac{1}{3}(\epsilon-4B)
\end{equation}
Despite its simplicity, since the equation of state is determined by only one parameter, into which all possible interactions have been absorbed, this model satisfactorily describes the basic properties of quark stars with an appropriate choice of the parameter $B$. In the present work is primarily used as a reference point for comparison with more elaborate models, such as the CFL model, presented below

\subsection{Quark stars: The Color-Flavor Locked  model}\label{CFL matter}

A considerable body of research has been devoted to  Witten’s proposal on the true ground state of matter \textcolor{blue}{\cite{Witten-1984}}, which was originally anticipated by Bodmer’s seminal work \textcolor{blue}{\cite{Bodmer-1971}}. According to this hypothesis, quark matter composed of $u$, $d$, and $s$ quarks, commonly referred to as strange matter, may possess an energy per baryon lower than that of both nuclear matter and $u$–$d$ quark matter. Together with the theoretical expectation of deconfined quark matter at supranuclear densities~\cite{Ivanenko-1965,Itoh-1970,Collins-1975,Weber-2005,schaffner-bielich_2020}, this idea has motivated extensive investigations into the existence of exotic compact objects known as strange stars~\cite{schaffner-bielich_2020,Glendenning-2000,Alock-1986,Hanesel-1986}. Due to their unusual composition, such stars are predicted to attain arbitrarily small masses and radii~\cite{schaffner-bielich_2020}.

At asymptotically high densities, the masses of quarks are negligibly small compared to the quark chemical potential. As is widely accepted under these conditions, strange quark matter most likely enters a superfluid phase, where quarks of all three flavors and colors form Cooper pairs and have the same Fermi momentum. This results in breaking the chiral symmetry and also prevents the presence of electrons. This phase is known as the Color-Flavor-Locked phase~\cite{Alford-2001a,
Alford-2008,Rajagopal-2000,Alford-1999,
oikonomou2023color,flores2017constraining}. Color-flavor locking affects significantly many features of quark matter, e.g. transport properties.

In total, self-bound stars that consist entirely of quark matter, from the center up to the
stellar surface, known as \textit{strange stars}, may occur for a wide range of parameters of the MIT bag model EOS \cite{lugones2002color}. Furthermore, research on the structure of these objects reveals that color superconductivity has a significant effect on the mass-radius relationship of strange stars, making it possible for very large maximum masses to appear \cite{lugones2003high,horvath2004self}.

Now, we are ready to present the thermodynamics of the CFL phase. The equation of state for CFL quark matter can be derived in the general framework of the MIT bag model. The pressure and energy density are given, in order of $\Delta^2$ ($\Delta$ is the gap parameter  representing the contribution of color superconductivity \cite{oikonomou2023color}) and $m_s^2$ (with $m_s$ the mass of the strange quark), as follows \cite{flores2017constraining}
\begin{eqnarray}\label{CFL_EOS_enrg2}
    \epsilon &=& 3P+4B-\frac{9\alpha\mu^2}{(\hbar c)^3\pi^2}, \nonumber \\  \mu^2&=&-3\alpha+\left[\frac{4}{3}\pi^2(B+P)(\hbar c)^3+9\alpha^2\right]^{1/2}  
\end{eqnarray}
where
\begin{equation}\label{alpha_cfl}
    \alpha=-\frac{m_s^2}{6}+\frac{2\Delta^2}{3}
\end{equation}
.


The absolute stability of the CFL quark matter requires the energy per baryon to be less than the neutron mass $m_n$ at vanishing pressure ($P=0$) and temperature ($T=0$). Thus, the following condition must be satisfied \cite{lugones2002color}
\begin{equation}\label{CFL_stable_baryonenrg}
    \frac{\epsilon}{n_B}\bigg|_{P=0}=3\mu\leq m_n = 939 \text{ MeV}
\end{equation}
This result is derived directly from the shared Fermi momentum among the three quark flavors in CFL matter and is valid at $T=0$ without any approximation. Since this condition must be fulfilled to the vanishing pressure points, using the second relation of Eq.~(\ref{CFL_EOS_enrg2}), we get \cite{flores2017constraining}
\begin{equation}\label{bag_high_constraint}
    B<-\frac{1}{(\hbar c)^3}\left(\frac{m_s^2m_n^2}{12\pi^2}+\frac{\Delta^2m_n^2}{3\pi^2}+\frac{m_n^4}{108\pi^2}\right)
\end{equation}
This equation defines a region in the $m_s-B$ plane in which
the energy per baryon is smaller than $m_n$ for a given $\Delta$. 

\subsection{The TOV equations }
The mechanical equilibrium of stellar matter is governed by a coupled system of two differential equations: the well-known Tolman–Oppenheimer–Volkoff  equations, together with the equation of state of the fluid, ${\cal \epsilon} = {\cal \epsilon}(P)$. This system can be expressed as follows~\cite{Shapiro:1983du,Haensel2007NeutronS1,schaffner-bielich_2020}
\begin{eqnarray}
\frac{dP(r)}{dr}&=&-\frac{G{\cal \epsilon}(r) M(r)}{c^2r^2}\left(1+\frac{P(r)}{{\cal \epsilon}(r)}\right) \nonumber \\
&\times&
 \left(1+\frac{4\pi P(r) r^3}{M(r)c^2}\right) \left(1-\frac{2GM(r)}{c^2r}\right)^{-1},
\label{TOV-1}
\end{eqnarray}
\begin{equation}
\frac{dM(r)}{dr}=\frac{4\pi r^2}{c^2}{\cal \epsilon}(r).
\label{TOV-2}
\end{equation}
The solution of the coupled differential equations (\ref{TOV-1}) and (\ref{TOV-2}) for $P(r)$ and $M(r)$ necessitates their numerical integration from the stellar center ($r = 0$) outward to the radius $r = R$, where the pressure effectively vanishes. At this point, the corresponding values of the stellar radius and gravitational mass are determined.

\section{Data analysis}
\subsection{Data generation}\label{data_gen}
For a given number $l$ of possible choices for $\Gamma_i$ and a certain number of $n$ polytropic segments, one can obtain 
\begin{equation}\label{number_of_polyEoSs}
    f=l^n
\end{equation}
differently parametrized hadronic EoSs. In this study, we assume the polytropic region extends over the interval [$\rho_0$, $7.5 \rho_0$], where $\rho_0=0.16 \ {\rm fm}^{-3}$ is the nuclear saturation density, which will also be denoted as $\rho_{\rm sat}$. We divide the region into four segments, with equal lengths in the logarithm of mass density $\rho$. For each segment, we allow four possible choices of $\Gamma_i$, namely $\{1,2,3,4\}$. We also employ HLPS-2 and HLPS-3, as the "main" EoSs before the transition to polytropic behavior~\cite{Hebeler-2013}. Thus, we result in $\rm f=256$ different mock EoSs per each "main" EoS and $512$ mock EoSs in total. 

\begin{figure}[h]
\centering
\includegraphics[width=240pt,height=12pc]{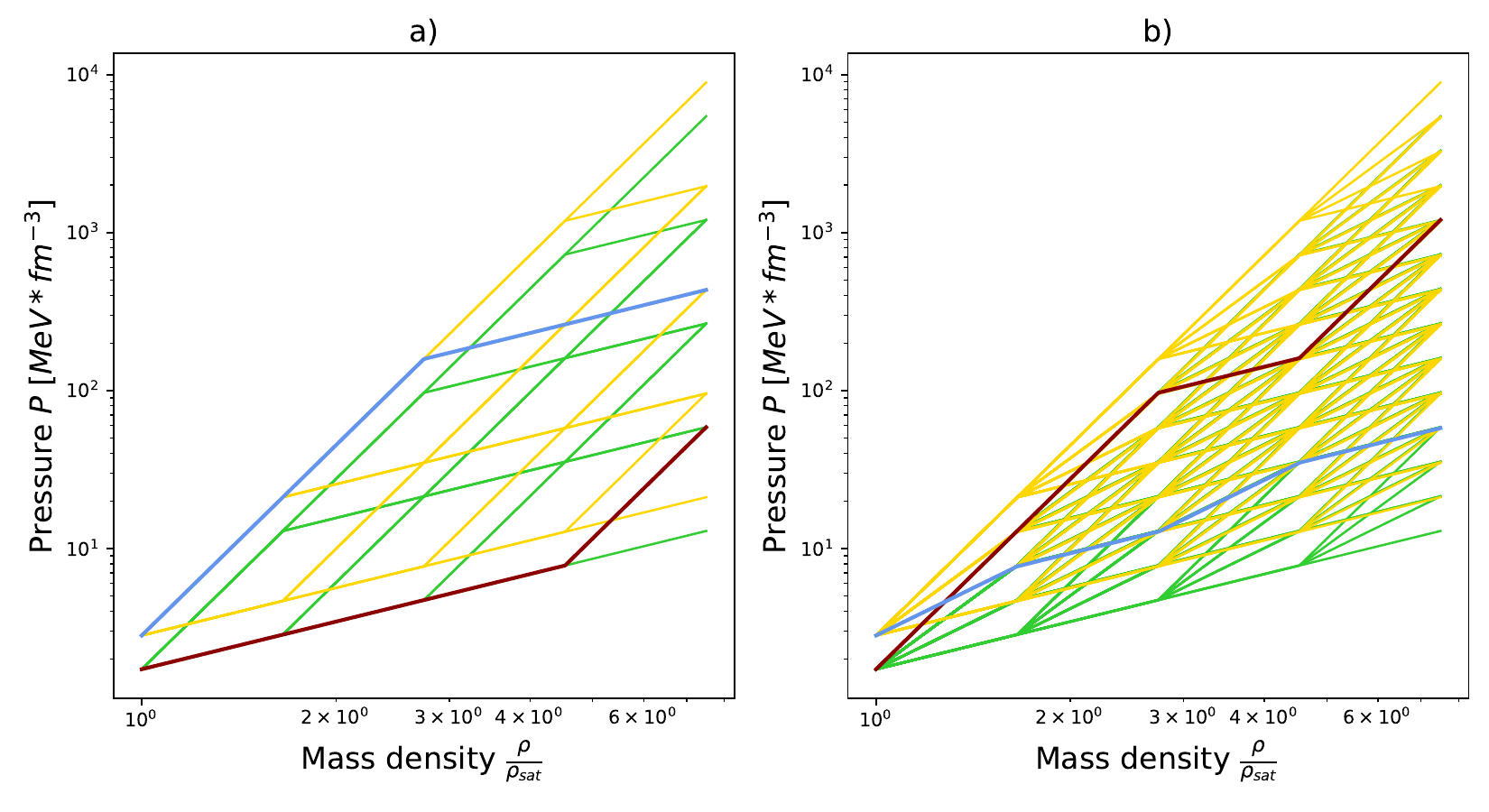}
\caption{Grids of polytropes for $n=4$ mass density segments and a) $l=2$ available choices $\{1,4\}$ or b) $l=4$ available choices $\{1,2,3,4\}$ for $\Gamma$ values. The green grid corresponds to polytropes that start at the nuclear saturation pressure of HLPS-2 ($1.722$ $ {\rm MeV\ fm}^{-3}$), while the yellow one to polytropes that start at the nuclear saturation pressure of HLPS-3 ($2.816$ $\rm MeV \ fm^{-3}$). The red line features (from left to right) the sequence $\{\Gamma:1\rightarrow{}1\rightarrow{}1\rightarrow{}4\}$ in a) and $\{\Gamma:4\rightarrow{}4\rightarrow{}1\rightarrow{}4\}$ in b). The blue line features (from left to right) the sequence $\{\Gamma:4\rightarrow{}4\rightarrow{}1\rightarrow{}1\}$ in a) and the sequence $\{\Gamma:2\rightarrow{}1\rightarrow{}2\rightarrow{}1\}$ in b).}
\label{NS_poly_grid}
\end{figure}

The corresponding grids are presented in FIG. \ref{NS_poly_grid}. It is evident that the option with four available values for $\Gamma_i$, offers a much more detailed scanning of the region, compared to the option with two values for $\Gamma_i$. Furthermore, in order to ensure that the majority of our mock EoSs reach the maximum mass point, we adjust the upper bound of the polytropic region $\rho_n$, after the division into segments. We do so based on the value of $\Gamma$  the first segment ($\Gamma_1$):
\begin{itemize}
    \item For $\Gamma_1\leq2$: $\rho_n=7.5\rho_0\rightarrow{15\rho_0}$ (soft).
    \item For $2<\Gamma_1\leq3$: $\rho_n=7.5\rho_0\rightarrow{12\rho_0}$ (intermediate).
    \item For $\Gamma_1>3$: $\rho_n=7.5\rho_0\rightarrow{9\rho_0}$ (stiff).
\end{itemize}

\begin{figure}[h]
\centering
\includegraphics[width=240pt,height=12pc]{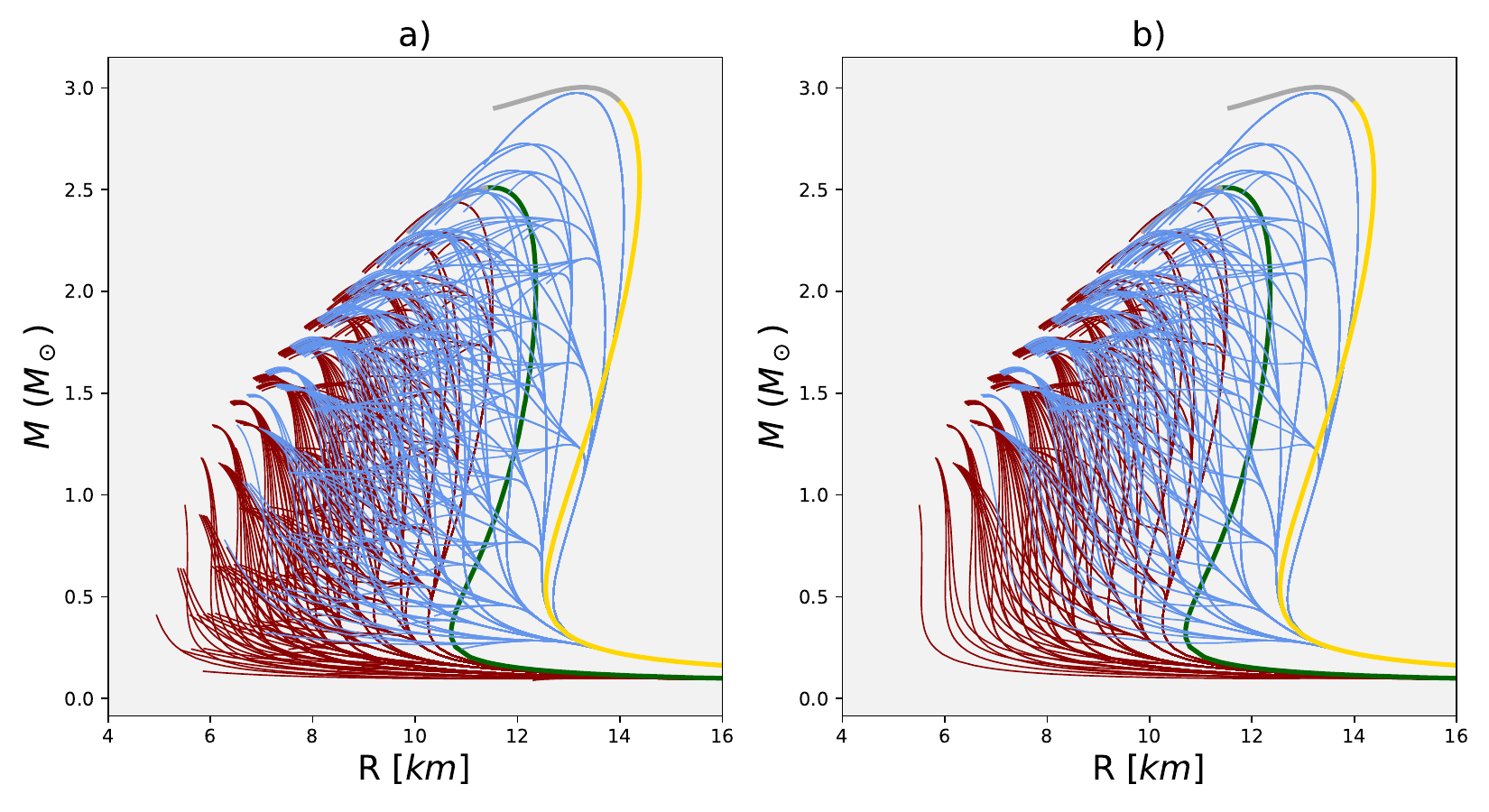}
\caption{Plots of the $M-R$ curves of mock Neutron Stars EoSs. a) The $M-R$ curves of all $512$ mock EoSs have derived from HLPS-2 and HLPS-3 'main' EoSs, for all $\Gamma$ combinations in four mass density segments. b) The $M-R$ curves of the $304$ out of $512$ mock EoSs, exceed the pressure of $850$ $\rm MeV\ fm^{-3}$. The $M-R$ curves of HLPS-2 (green) and HLPS-3 (yellow) are also included, using gray endings to mark the violation of causality.}
\label{NS_poly_MR}
\end{figure}

The resulting $\rm M-R$ curves are presented in Fig.~\ref{NS_poly_MR}. Now, in our analysis, we aim to reconstruct an EoS by predicting values of the energy density $\epsilon$ in the pressure range $[10,800]$ $\rm MeV\ fm^{-3}$. Of course, this means that all the mock EoSs, that do not exceed this pressure value, need to be ignored. For more acceptable results, an extra threshold of $50$ ${\rm MeV}\ {\rm fm}^{-3}$ is imposed and we choose to neglect all the EoSs that terminate under the pressure of $850$ ${\rm MeV}\ {\rm fm}^{-3}$. This filtering leaves $304$ mock EoSs to work with, as shown in b) of Fig.~ \ref{NS_poly_MR}, scanning the $\rm M-R$ space equally well as the $512$ ones. 

These EoSs are labeled as: {\bf HLPS-X\_$\Gamma_1\Gamma_2\Gamma_3\Gamma_4$L}, where X takes the value of $2$ or $3$, based on the "main" EoS that was used to produce the mock polytropic EoS. The values of $\Gamma_i$ are included in a coded format. Since we considered $4$ possible choices for each $\Gamma_i$, we assign a letter to each of them, starting from $\rm A$. A: $\Gamma_i=1$, B: $\Gamma_i=2$, C: $\Gamma_i=3$ and D: $\Gamma_i=4$. The suffix $\rm L$ indicates that the possible violation of causality has been taken into account and handled accordingly, by including a linear part to the EoS (see \ref{PolyLin_EOSs}).

The methodology employed to generate data for quark stars differs significantly. One simply has to change the values of the parameters $\rm B$ and $\rm \Delta$, in order to obtain a large amount of MIT bag or CFL EoSs. Regarding the MIT bag model, we assume that parameter $\rm B$  takes values in the interval $[60,250]$ and uses a step of $0.5$ $\rm MeV\ fm^{-3}$. Following this way, we generate $381$ EoSs, representing the MIT bag model (labeled as: {\bf MITbag-X, X:1,..,381}).

\begin{figure}[h]
\centering
\includegraphics[width=240pt,height=12pc]{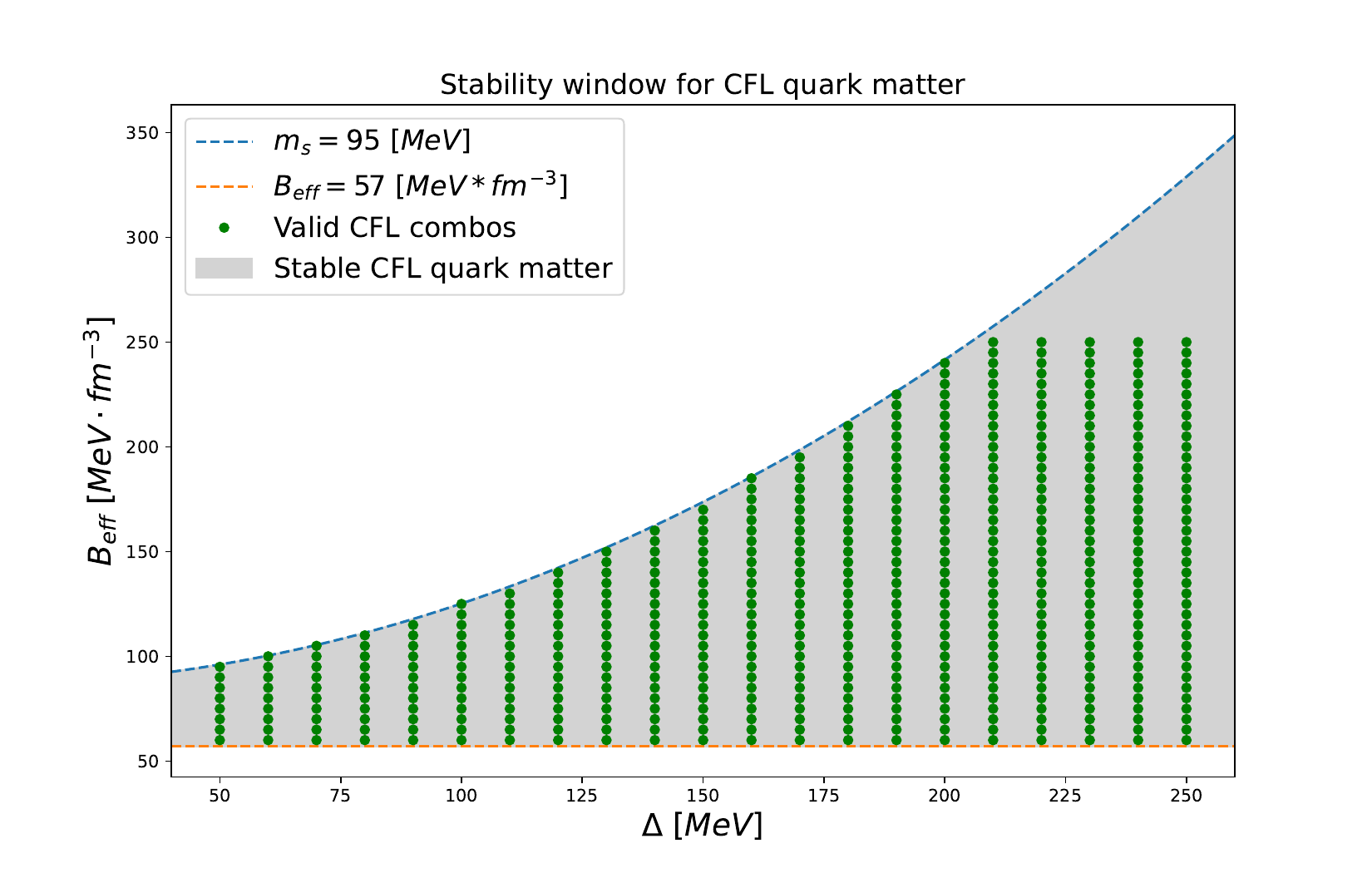}
\caption{Scanning the stability window region for CFL quark matter with $m_s=95$ $\rm MeV$. Keeping $m_s$ value constant, Eq. \ref{bag_high_constraint} yields to the equation of a curve, namely the $B_{max}=B_{max}(\Delta)$ curve, as shown with blue color and dashed-line style. The orange dashed-line curve marks the minimum value of $57$ $\rm MeV\cdot fm^{-3}$. The coordinates of the green points, which scan the stability region, are valid combinations of $B$ and $\Delta$ values, for $m_s=95$ $\rm MeV$.}
\label{QS_valid_cfl}
\end{figure}

In contrast, the constraints applied in the CFL model (see \ref{CFL matter}), form a stability window in the $\rm B-\Delta$ space, as shown in gray Fig.~\ref{QS_valid_cfl}. Assuming that $B$ takes values in the interval $[60,250]$ $\rm MeV\ fm^{-3}$, with a step of $5$ $\rm MeV\ fm^{-3}$ and $\Delta$ takes values in the interval $[50,250]$ $\rm MeV$, with a step of $10$ $\rm MeV$, leads to $510$ different valid EoSs, representing the CFL model (labeled as: {\bf CFL-Y, Y:1,...,510}).

\begin{figure}[h]
\centering
\includegraphics[width=240pt,height=12pc]{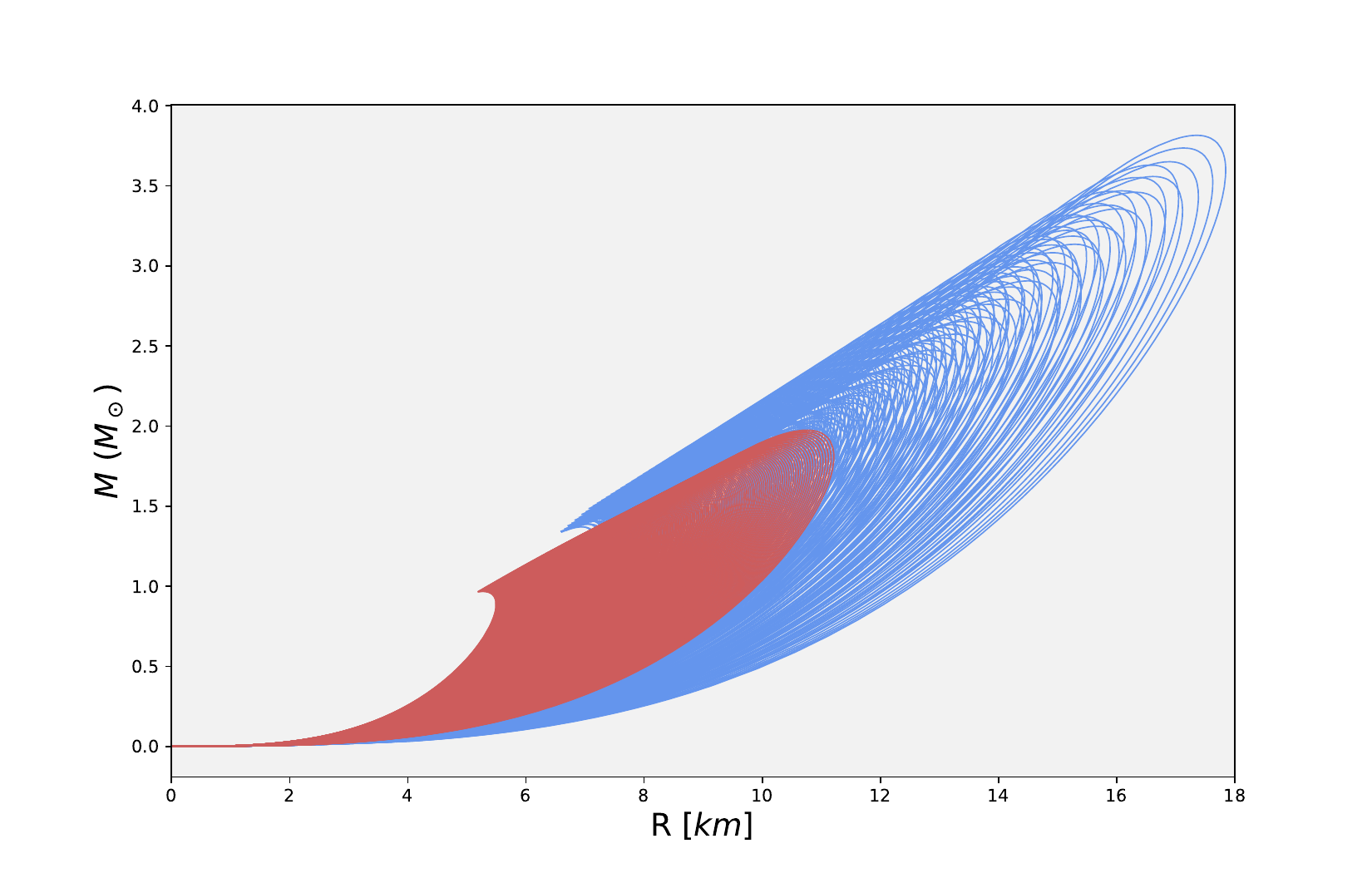}
\caption{Plots of the $M-R$ curves of Quark Stars EoSs. The $381$ $M-R$ curves of MIT bag model EoSs are shown in red, while the $510$ $M-R$ curves of CFL model EoSs are shown in blue.}
\label{QS_MR_all}
\end{figure}

Accordingly, we obtain 891 quark-matter curves spanning the M–R plane, as shown in Fig.~\ref{QS_MR_all}.  All these curves start at very small masses for small radii, as expected for self-bound stars without crust. However, the CFL ones (blue) extend over a much wider range in the $M-R$ plane, reaching masses nearly four times the mass of the Sun ($M\sim 4M_\odot$) and radii greater than $16\rm km$ ($R\geq16\rm km$), in some extreme cases. The overlap between the curves of the two models is also, present and covers a fairly large area: $M\leq2M_\odot$ and $R\leq11\rm km$.

\subsection{Data preparation}\label{data_prep}
\begin{figure}[h]
\centering
\includegraphics[width=240pt,height=13pc]{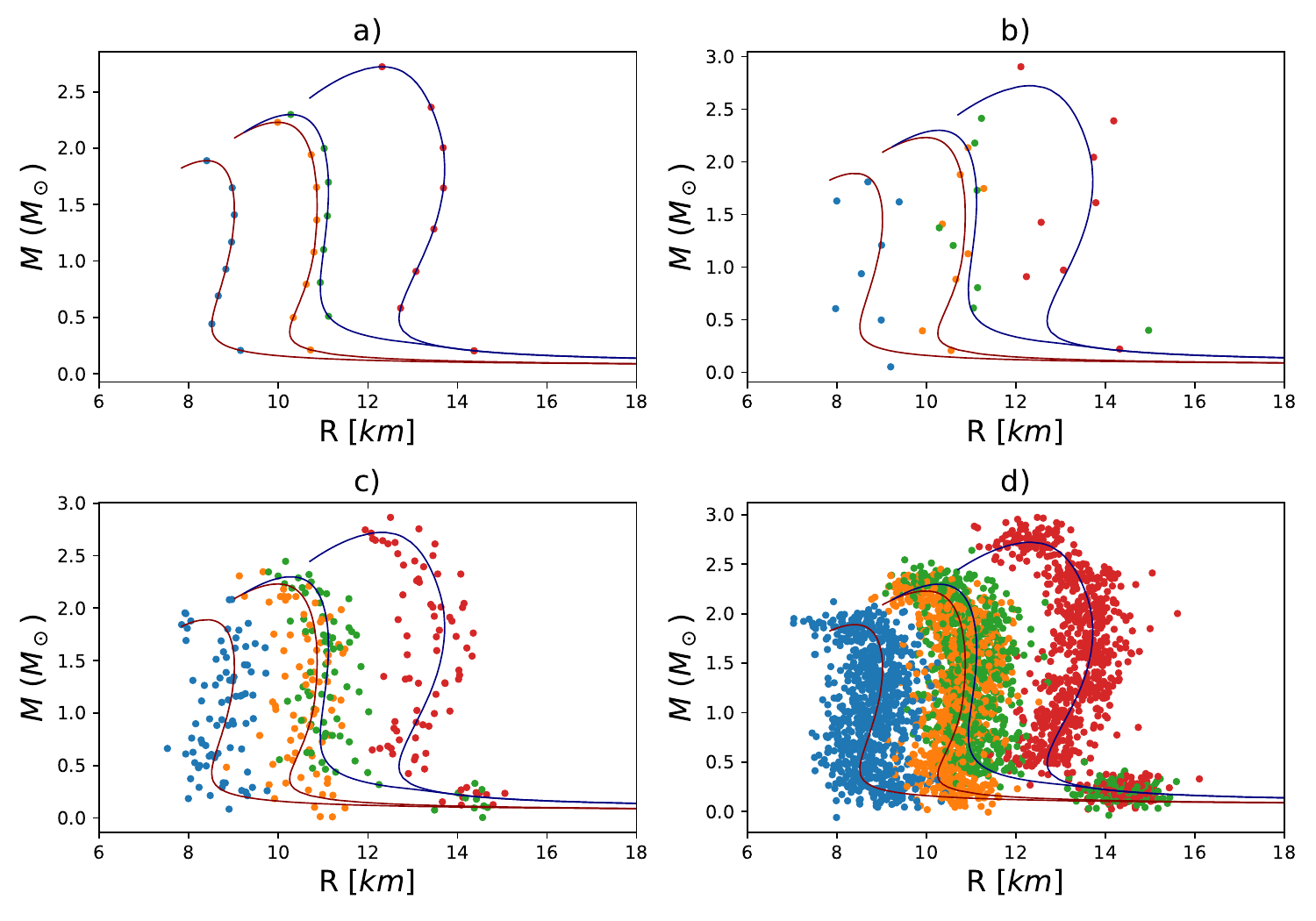}
\caption{Sampling example of mass and radius data, using $8$ points from each of the M-R curves of the following polytropic EoSs: HLPS-2\_ADDDL (blue), HLPS-2\_DCDCL (orange),  HLPS-3\_ADDDL (green) and HLPS-3\_DCDCL (red). The respective M-R curves are plotted too. The graphs depict: a) the noise-free basic observation of M-R points for each EoS, b) $1$ random M-R observation per EoS, c) $10$ random M-R observations per EoS and d) $100$ random M-R observations per EoS. Each random observation includes additional observational noise: $\Delta M\sim0.1M_\odot$ and $\Delta R\sim0.5 \ {\rm km}$.}
\label{NS_sample_MR}
\end{figure}

\begin{figure}[h]
\centering
\includegraphics[width=240pt,height=13pc]{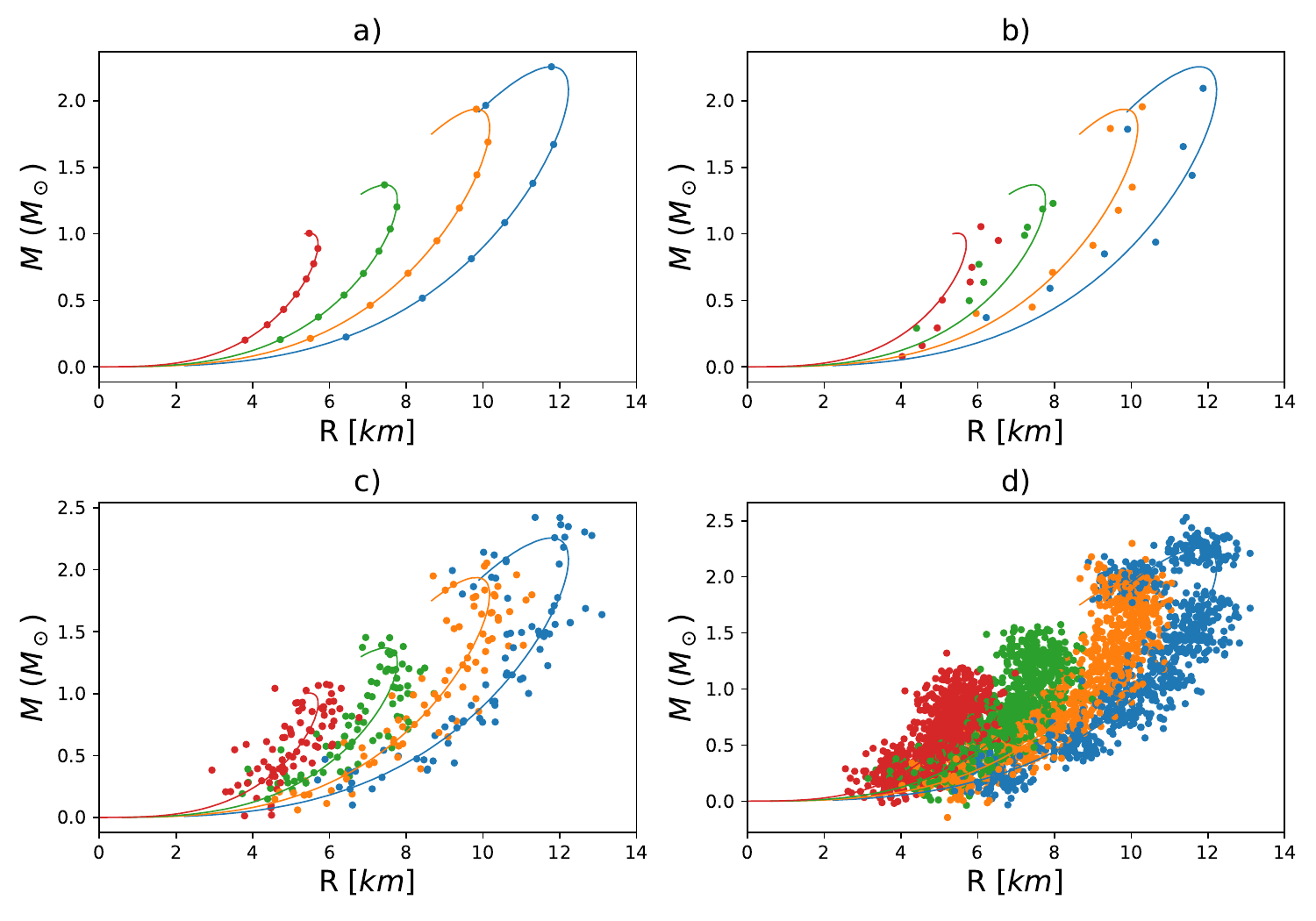}
\caption{Sampling example of mass and radius data, using $8$ points from each of the M-R curves of the following quark matter EoSs: CFL-50 (blue), CFL-250 (orange), MITbag-131 (green) and MITbag-345 (red). The respective M-R curves are plotted too. The graphs depict: a) the noise-free basic observation of M-R points for each EoS, b) $1$ random M-R observation per EoS derived, c) $10$ random M-R observations per EoS and d) $100$ random M-R observations per EoS. Each random observation includes additional observational noise: $\Delta M\sim0.1M_\odot$ and $\Delta R\sim0.5km$.}
\label{QS_sample_MR}
\end{figure}

\begin{figure}[h]
\centering
\includegraphics[width=240pt,height=13pc]{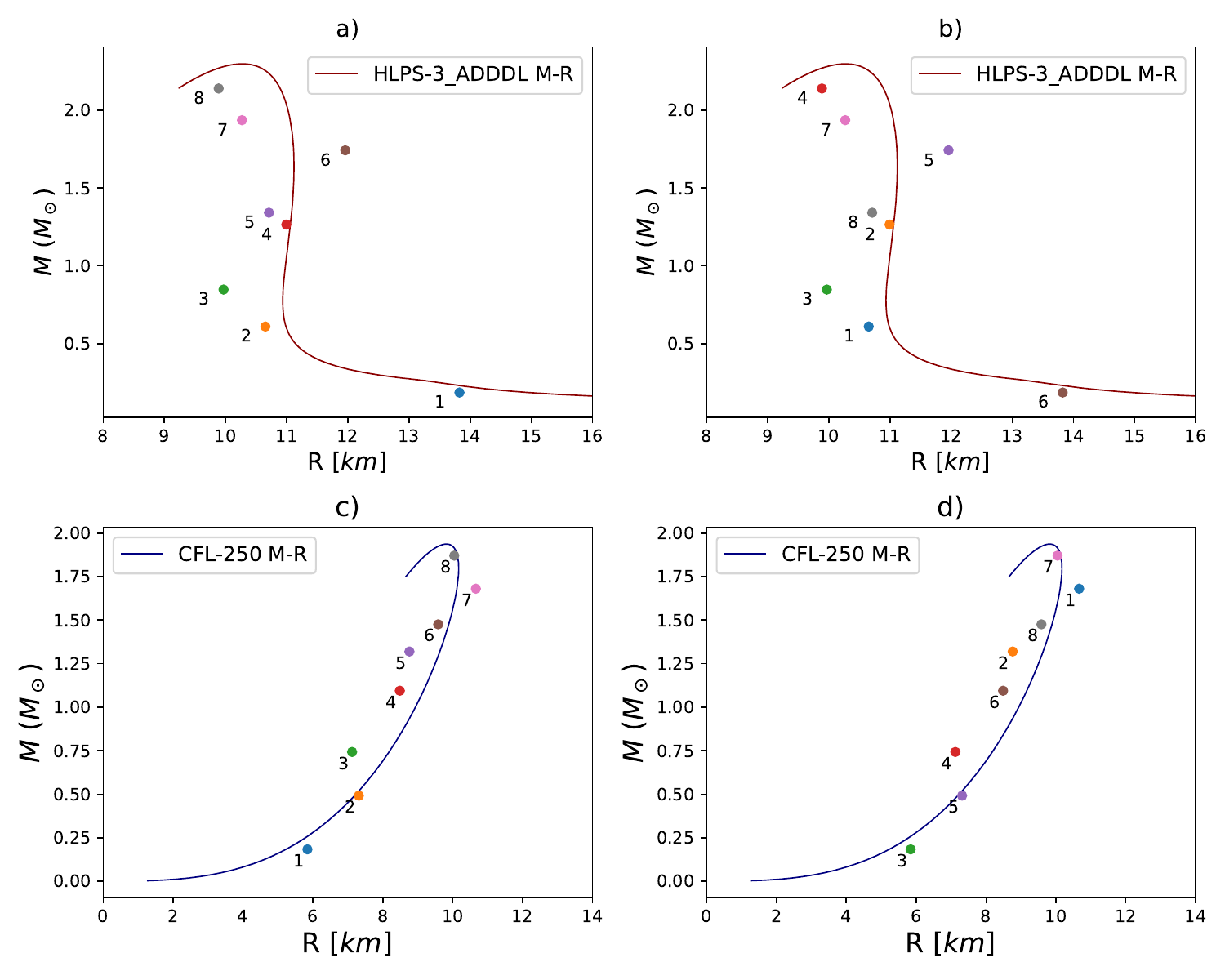}
\caption{Shuffling example of M-R data from a) and b) a neutron star M-R curve, c) and d) a quark star M-R curve. The numbers indicate the order in which the points will be recorded in the final dataset. Thus, without shuffling, the points are recorded in ascending order of mass, as shown in a) and c). On the contrary, after shuffling, the points are recorded in random order, as shown in b) and d). Notice, that the points include observational noise and that the shuffling affects only the order of recording and not their coordinates.}
\label{shuffle_MR}
\end{figure}

Having obtained the $M-R$ curves from the solution of TOV equations, we continue with the sampling of our data. From each $M-R$ curve, we get either $8$ or $16$ points, with masses equally spaced in the interval $[0.2M_\odot,M_{max}]$. These points are our initial, noise-free observation: $(M^{(0)}_j,R^{(0)}_j)$ of the curve. Next, we add normally distributed errors in order to artificially simulate observational noise. We choose errors with standard deviation: $\Delta M=0.1M_{\odot}$ and $\Delta R=0.5 {\rm km}$, for the mass and radius values, respectively. By adding these errors to the points of the initial observation, we can obtain unique random observations: $(M^{(i)}_j,R^{(i)}_j)$. We found that 100 random observations satisfactorily represent an $M-R$ curve shown in Fig.~\ref{NS_sample_MR} and Fig.~\ref{QS_sample_MR}, following previous studies \cite{Fujimoto-2018}. Hence, we result in $30.400$ rows of neutron star data and $89.100$ rows of quark star data. We also shuffle $M-R$ pairs when recording random observations, in order to avoid possible linear correlations between predictors. An example of shuffling is shown in Fig.~\ref{shuffle_MR}.

Regarding the target variables, we sample the energy density values of  $\rm \epsilon$ at 12 
pressure values. For the polytropic Neutron Star EoSs, we choose these pressure values to be: \{10, 25, 50, 75, 100, 200, 300, 400, 500, 600, 700, 800\} $\rm MeV\ fm^{-3}$. Notice that the sampling is denser in the region [10,100] ${\rm MeV}\ {\rm fm}^{-3}$. In this region, the hadronic EoSs turn stiffer and more points need to be added for detailed capture of the EoS's behavior. In the other region, [100, 800] $\rm MeV\ fm^{-3}$, the pressure points are equally spaced. The final value $P_{12}=800$ $\rm MeV\ fm^{-3}$, is arbitrarily selected, in order to reach sufficiently large values of mass density and exceed (in most cases) the maximum mass point. Following the same approach for quark star EoSs, we choose the $12$ pressure values to be: \{10, 100, 200, 300, 400, 500, 600, 700, 800, 900, 1000, 1100\} $\rm MeV\ fm^{-3}$. The sampling is, this time, uniformly distributed in terms of pressure axis, since the quark matter EoSs are linear (or almost linear) and do not exhibit sudden changes in their behavior. Thus, no denser sampling is needed in some pressure regions. This allows us to increase the final sampling pressure $P_{12}$, from 800 to 1100 $\rm MeV\ fm^{-3}$, and be able to reconstruct the quark matter EoSs in a larger pressure region.

\subsection{Machine learning and deep learning algorithms}\label{fine_tune}
We employed several machine-learning algorithms,  including  Decision Trees, Random Forests, Gradient Boosting, XGBoost, and Deep Learning (Artificial Neural Networks), to address the problem using different model classes. Their Python implementations natively support multivariate multiple regression. Additionally, we utilized 5-fold cross-validation and grid search to optimize model performance (see hyperparameter TABLES \ref{tab:tune_dtree},\ref{tab:tune_rf}
\ref{tab:tune_gradboost},\ref{tab:tune_xgboost}). 

\textbf{Decision Trees.}  
Regression trees \cite{CART} partition the data to minimize target variance within subsets, predicting by the mean at the terminal leaves. They capture nonlinear relations but are prone to overfitting. See TABLE \ref{tab:tune_dtree}.

\textbf{Random Forests.}  
Random Forests \cite{Breiman2001} average predictions from multiple bootstrap-trained trees, improving accuracy and reducing overfitting at the cost of higher computation. See TABLE \ref{tab:tune_rf}.

\textbf{Gradient Boosting.}  
Gradient Boosting regression \cite{Friedman2001, Natekin2013} sequentially builds an ensemble of weak learners, each trained to correct the residual errors of its predecessors. This method achieves high predictive performance but can also be prone to overfitting if not carefully regularized. See TABLE \ref{tab:tune_gradboost}.

\textbf{XGBoost.}  
Extreme Gradient Boosting (XGBoost) \cite{Chen2016}is an optimized form of gradient boosting that iteratively fits residuals while incorporating regularization and computational enhancements. It offers high accuracy and efficiency, though often at the expense of interpretability. See TABLE \ref{tab:tune_xgboost}.

\textbf{Deep Learning (Artificial Neural Networks).}  
Artificial neural networks (ANNs) \cite{Hornik1989} can also be used for regression. They are especially effective when large datasets are available, enabling the learning of highly nonlinear and complex feature--target relationships. Input features are fed into the input layer, propagated through one or more hidden layers with nonlinear activation functions, and finally mapped to outputs in the last layer, which matches the dimensionality of the target variables. See TABLE \ref{tab:DNN_structure}. 

 These methods were chosen because they combine predictive accuracy, flexibility, 
 and well-established implementations in Python programming language that support multivariate multiple 
 regression. \textsc{Decision Trees} provide interpretability and a baseline model, 
 while ensemble methods such as \textsc{Random Forests} and \textsc{Gradient Boosting} 
 reduce variance and bias, achieving higher accuracy. \textsc{XGBoost} extends these 
 ideas with regularization and computational optimizations, making it one of the most 
 efficient boosting algorithms in practice. Finally, \textsc{Deep Learning} (ANNs) 
 is included as it can capture highly nonlinear relationships when sufficient data are 
 available. Together, these algorithms span a spectrum from interpretable to highly 
 expressive models, allowing us to balance performance, generalization, and complexity.

\begin{table}[h]
    \centering
    \begin{tabular}{|p{3cm}|p{2.5cm}|p{2.5cm}|}
    \hline
    \hline
    \textbf{Hyperparameters}  &  \textbf{Values NS} & \textbf{Values QS} \\
    \hline
    \hline
    \vspace{0.005cm} max\_depth   & \vspace{0.005cm} [None, 5, 10, \textcolor{green!60!black}{20}] & \vspace{0.005cm} [None, 5, 10, \textcolor{green!60!black}{20}] \\
    \hline
    \vspace{0.005cm} max\_features   & \vspace{0.005cm} [\textcolor{green!60!black}{None}, 'sqrt', 'log2'] & \vspace{0.005cm} [\textcolor{red}{None}, 'sqrt', \textcolor{blue}{'log2'}] \\
    \hline
    \vspace{0.005cm} criterion  & \vspace{0.005cm} ['squared\_error', \textcolor{green!60!black}{'friedman\_mse'}] & \vspace{0.005cm} ['squared\_error', \textcolor{green!60!black}{'friedman\_mse'}] \\
    \hline 
    \hline
    \end{tabular}
    \caption{Values of most significant hyperparameters for the tuning of \textit{Decision Tree} models. Applying {\bf Grid Search}, the resulted values for $16$ features are highlighted with \textcolor{red}{red}, while the resulted values for $32$ features are highlighted with \textcolor{blue}{blue}. If the values are same for $16$ and $32$ features, they are shown in \textcolor{green!60!black}{green}.}
    \label{tab:tune_dtree}
\end{table}

\begin{table}[h]
    \centering
    \begin{tabular}{|p{3cm}|p{2.5cm}|p{2.5cm}|}
    \hline
    \hline
    \textbf{Hyperparameters}  &  \textbf{Values NS} & \textbf{Values QS} \\
    \hline
    \hline
    \vspace{0.005cm} n\_estimators  & \vspace{0.005cm} [25, \textcolor{green!60!black}{50}] & \vspace{0.005cm} [25, \textcolor{green!60!black}{50}] \\
    \hline
    \vspace{0.005cm} max\_depth   & \vspace{0.005cm} [\textcolor{green!60!black}{None}, 10, 20] & \vspace{0.005cm} [\textcolor{green!60!black}{None}, 10, 20] \\
    \hline
    \vspace{0.005cm} max\_features   & \vspace{0.005cm} [\textcolor{green!60!black}{None}, 'sqrt', 'log2'] & \vspace{0.005cm} [\textcolor{green!60!black}{None}, 'sqrt', 'log2'] \\
    \hline
    \vspace{0.005cm} criterion  & \vspace{0.005cm} [\textcolor{green!60!black}{'squared\_error'}] & \vspace{0.005cm} [\textcolor{green!60!black}{'squared\_error'}] \\
    \hline 
    \hline
    \end{tabular}
    \caption{Values of most significant hyperparameters for the tuning of \textit{Random Forest} models. Applying {\bf Grid Search}, the resulted values for $16$ features are highlighted with \textcolor{red}{red}, while the resulted values for $32$ features are highlighted with \textcolor{blue}{blue}. If the values are same for $16$ and $32$ features, they are shown with \textcolor{green!60!black}{green}.}
    \label{tab:tune_rf}
\end{table}

\begin{table}[h!]
    \centering
    \begin{tabular}{|p{3cm}|p{2.5cm}|p{2.5cm}|}
    \hline
    \hline
    \textbf{Hyperparameters}  &  \textbf{Values NS} & \textbf{Values QS} \\
    \hline
    \hline
    \vspace{0.005cm} n\_estimators  & \vspace{0.005cm} [50, \textcolor{green!60!black}{100}] & \vspace{0.005cm} [50, \textcolor{green!60!black}{100}] \\
    \hline
    \vspace{0.005cm} learning\_rate  & \vspace{0.005cm} [0.01, \textcolor{green!60!black}{0.05}] & \vspace{0.005cm} [0.01, \textcolor{green!60!black}{0.05}] \\
    \hline
    \vspace{0.005cm} max\_depth   & \vspace{0.005cm} [3, \textcolor{green!60!black}{5}] & \vspace{0.005cm} [3, \textcolor{green!60!black}{5}] \\
    \hline
    \hline
    \vspace{0.005cm} max\_features   & \vspace{0.005cm} [\textcolor{green!60!black}{'sqrt'}, 'log2'] & \vspace{0.005cm} [\textcolor{green!60!black}{'sqrt'}, 'log2'] \\
    \hline 
    \vspace{0.005cm} loss  & \vspace{0.005cm} [\textcolor{green!60!black}{'squared\_error'}] & \vspace{0.005cm} [\textcolor{green!60!black}{'squared\_error'}] \\
    \hline 
    \hline
    \end{tabular}
    \caption{Values of most significant hyperparameters for the tuning of \textit{Gradient Boosting} models. The resulted values for $16$ features are highlighted with \textcolor{red}{red}, while the resulted values for $32$ features are highlighted with \textcolor{blue}{blue}. If the values are same for $16$ and $32$ features, they are shown with \textcolor{green!60!black}{green}.}
    \label{tab:tune_gradboost}
\end{table}

\begin{table}[h!]
    \centering
    \begin{tabular}{|p{3cm}|p{2.5cm}|p{2.5cm}|}
    \hline
    \hline
    \textbf{Hyperparameters}  &  \textbf{Values NS} & \textbf{Values QS} \\
    \hline
    \hline
    \vspace{0.005cm} n\_estimators  & \vspace{0.005cm} [50, \textcolor{green!60!black}{100}] & \vspace{0.005cm} [50, \textcolor{green!60!black}{100}] \\
    \hline
    \vspace{0.005cm} learning\_rate  & \vspace{0.005cm} [0.05, \textcolor{green!60!black}{0.1}] & \vspace{0.005cm} [0.05, \textcolor{green!60!black}{0.1}] \\
    \hline
    \vspace{0.005cm} subsample  & \vspace{0.005cm} [0.7, \textcolor{green!60!black}{1.0}] & \vspace{0.005cm} [\textcolor{green!60!black}{0.7}, 1.0] \\
    \hline
    \vspace{0.005cm} max\_depth   & \vspace{0.005cm} [3, 5, \textcolor{green!60!black}{7}] & \vspace{0.005cm} [3, 5, \textcolor{green!60!black}{7}] \\
    \hline 
    \hline
    \end{tabular}
    \caption{Values of most significant hyperparameters for the tuning of \textit{XGBoost} models. The resulted values for $16$ features are highlighted with \textcolor{red}{red}, while the resulted values for $32$ features are highlighted with \textcolor{blue}{blue}. If the values are same for $16$ and $32$ features, they are shown with \textcolor{green!60!black}{green}.}
    \label{tab:tune_xgboost}
\end{table}

\begin{table}[h!]
    \centering
    \begin{tabular}{|p{3cm}|p{3cm}|p{2cm}|}
    \hline
    \hline
    \textbf{Layers} & \textbf{Neuron Numbers} &  \textbf{Activation}\\
    \hline
    \hline
    Dense\_0 (Input) & 16 or 32, based on features & - \\
    \hline
    Dense\_1 & 128 & ReLU \\
    Batch\_Normalization & 128 & - \\
    Dropout & drops $50\%$ of neurons & - \\
    \hline
    Dense\_2 & 64 & ReLU \\
    Batch\_Normalization & 64 & - \\
    Dropout & drops $50\%$ of neurons & - \\
    \hline
    Dense\_3 & 32 & ReLU \\
    Batch\_Normalization & 32 & - \\
    \hline
    Dense\_4 (Output) & 12 & - \\
    \hline
    \hline
    \end{tabular}
    \caption{The structure of the four (two for quark stars and two for neutron stars, of 16 and 32 features respectively) Deep Neural Network (\textit{DNN}) models we built for the purposes of this work. Each \textit{DNN} features 3 hidden layers and incorporates the techniques of batch normalization and dropout to prevent overfitting. We selected the name \textit{DNN-3} for these models. }
    \label{tab:DNN_structure}
\end{table}




\section{Results and Discussion}

\begin{figure}[h]
\centering
\includegraphics[width=240pt,height=13pc]{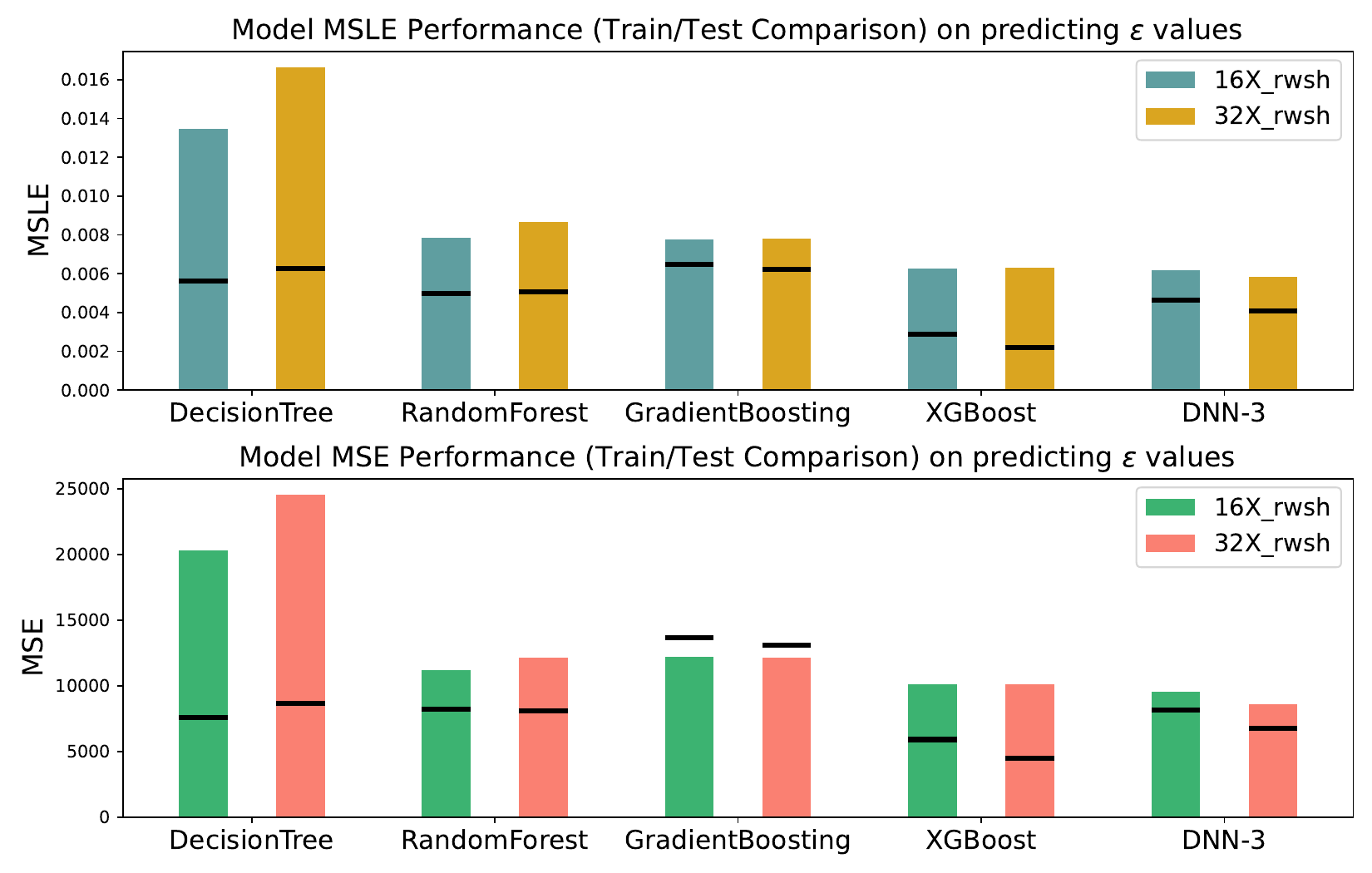}
\caption{Metric results for regression on Neutron Stars data. \textit{Top}: MSLE results. \textit{Bottom}: MSE results. The results are presented in grouped bar plots. Each group contains two bar plots, one for each distinct number of features (\textit{left bar}: 16 features, \textit{right bar}: 32 features). The black lines in each bar, depict the respective training result, i.e. the performance of the model on the training dataset itself, after fitting. }
\label{metrics_results_NS}
\end{figure}

\begin{figure}[h]
\centering
\includegraphics[width=240pt,height=13pc]{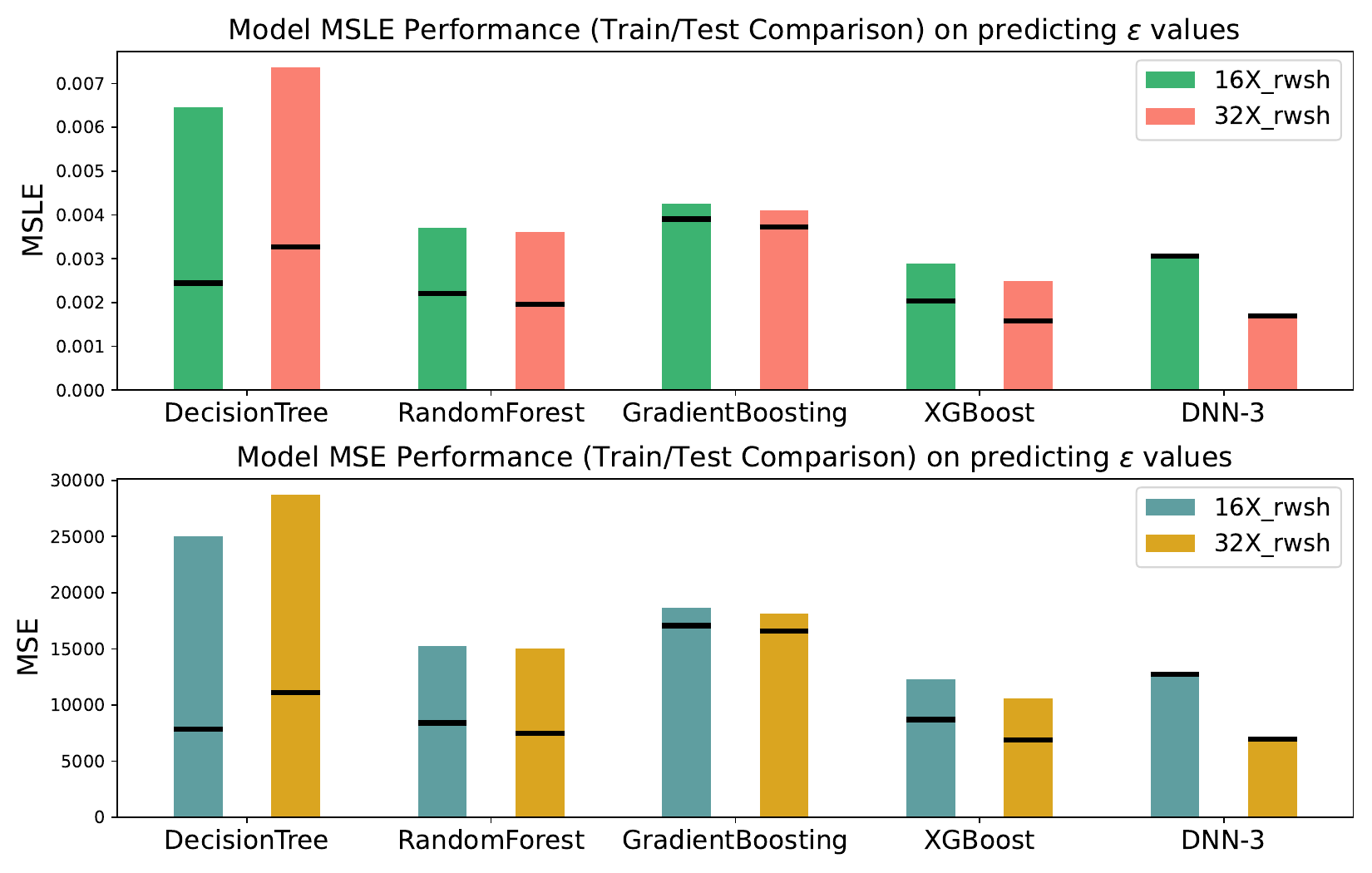}
\caption{Metric results for regression on Quark Stars data. \textit{Top}: MSLE results. \textit{Bottom}: MSE results. The results are presented in grouped bar plots. There are as many groups as the different algorithms used. Each group contains two bar plots, one for each distinct number of features (\textit{left bar}: 16 features, \textit{right bar}: 32 features). The black lines in each bar, depict the respective training result, i.e. the performance of the model on the training dataset itself, after fitting.}
\label{metrics_results_QS}
\end{figure}


After fitting the regression models, their performance is evaluated for the training dataset and the test dataset to calculate the accuracy and check for possible overfitting. An overview of our results for regression on neutron star data is presented in Fig.~\ref{metrics_results_NS}, where two metric functions are used: MSLE and MSE. At first glance, one can notice the weaker performance of the \textit{Decision Tree} models, compared to all other machine learning models. On the other hand, based exclusively on the test results of the metrics, we can conclude that the \textit{XGBoost} models are the best-performing machine learning models in our study, as they compete closely with the more sophisticated \textit{DNN-3} models. An important notice, concerning all models, is the fact that their accuracy is similar, regardless of the number of features used in the input, as can be seen in Fig.~\ref{metrics_results_NS}. A similar conclusion can be drawn regarding the performance of our regression models on the quark star dataset, shown in Fig.~\ref{metrics_results_QS}.

\begin{figure*}[htb]
\centering
\includegraphics[width=0.45\textwidth, height=4.5cm]{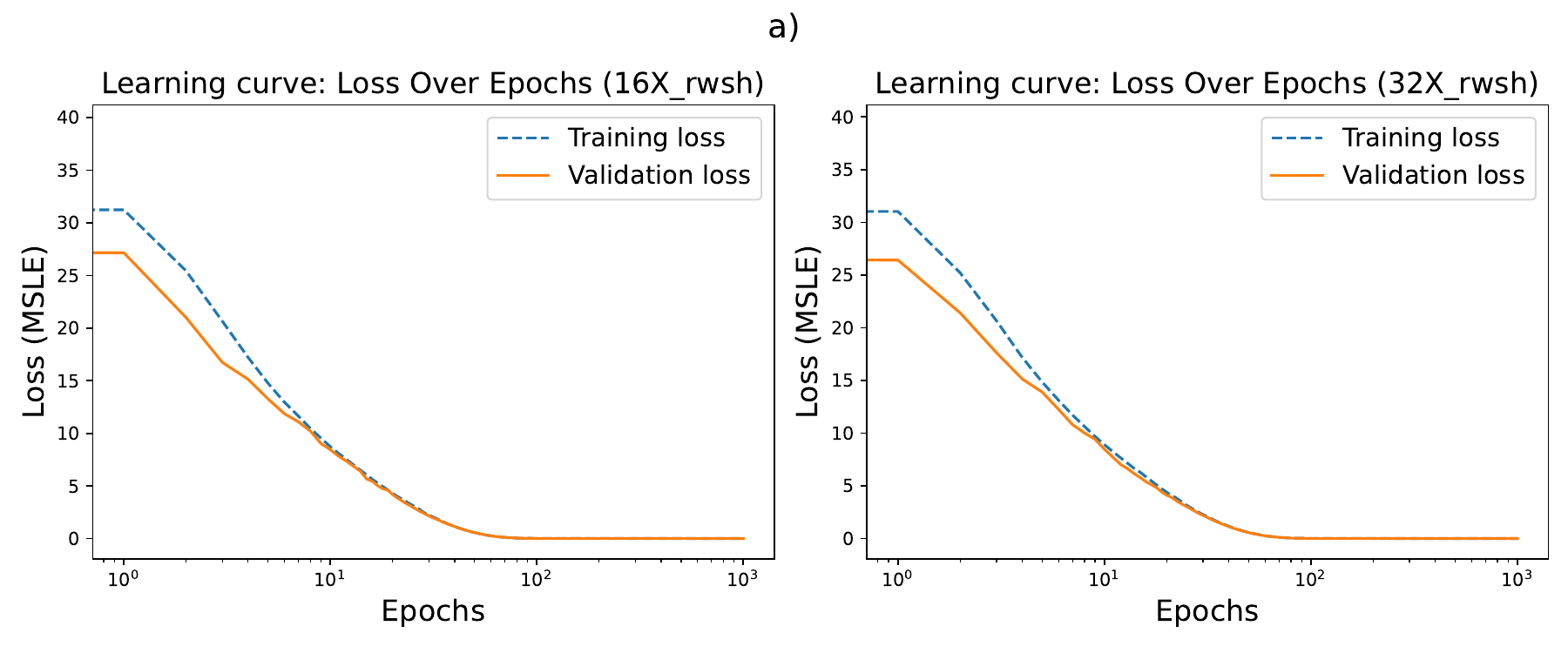}
\includegraphics[width=0.45\textwidth, height=4.5cm]{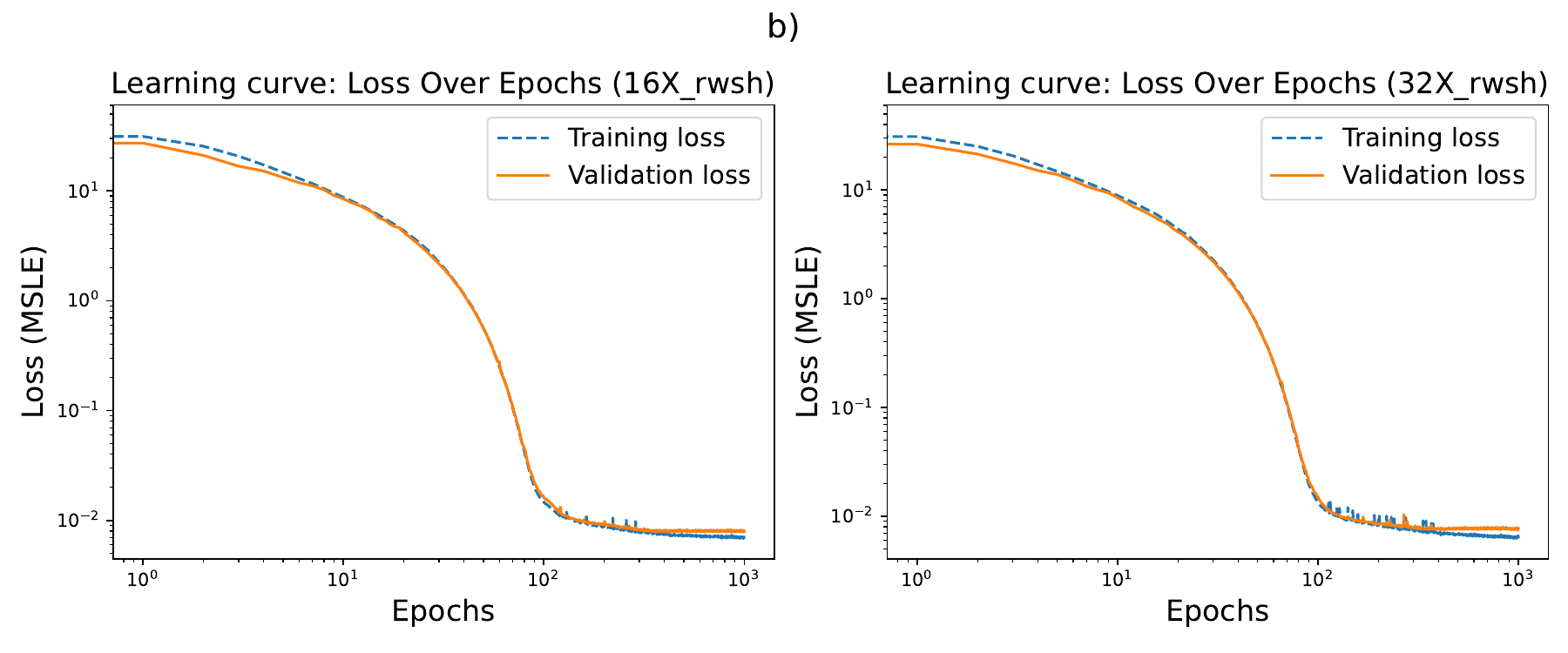}
\includegraphics[width=0.45\textwidth, height=4.5cm]{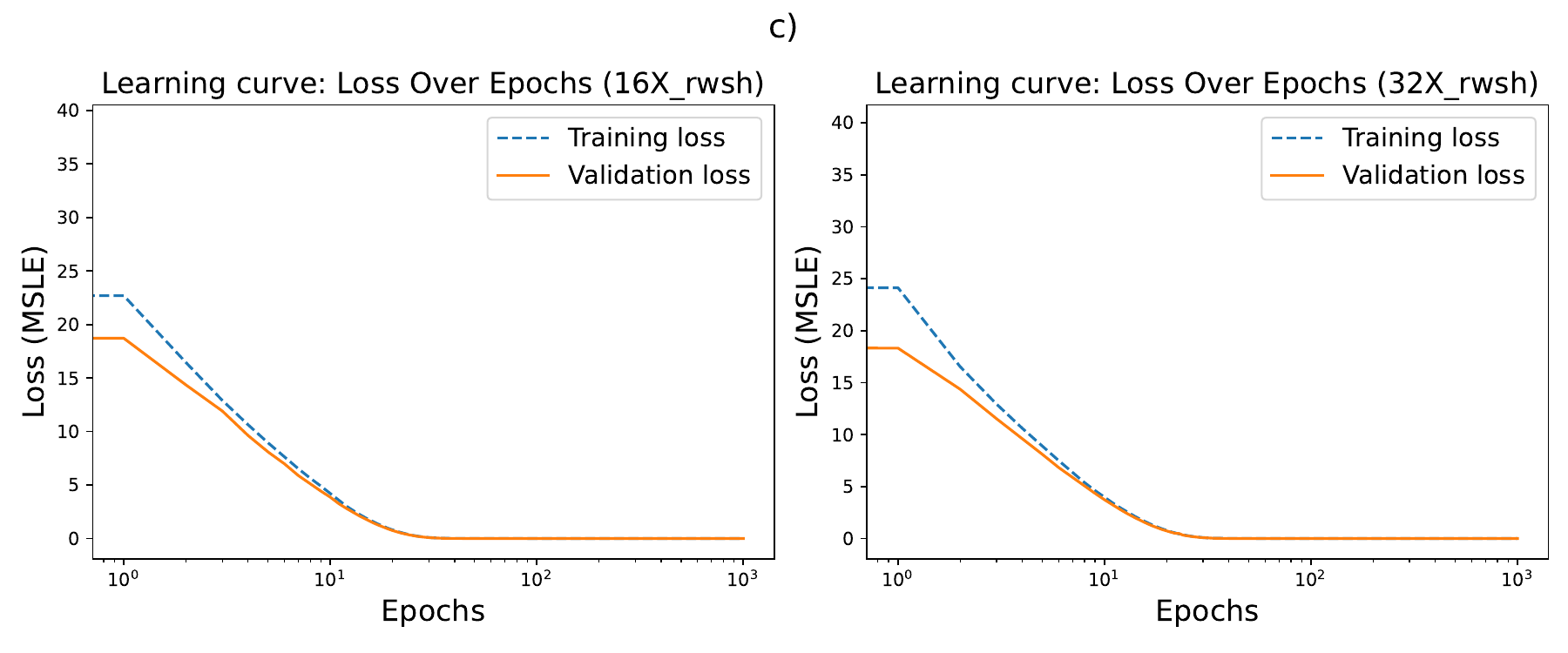}
\includegraphics[width=0.45\textwidth, height=4.5cm]{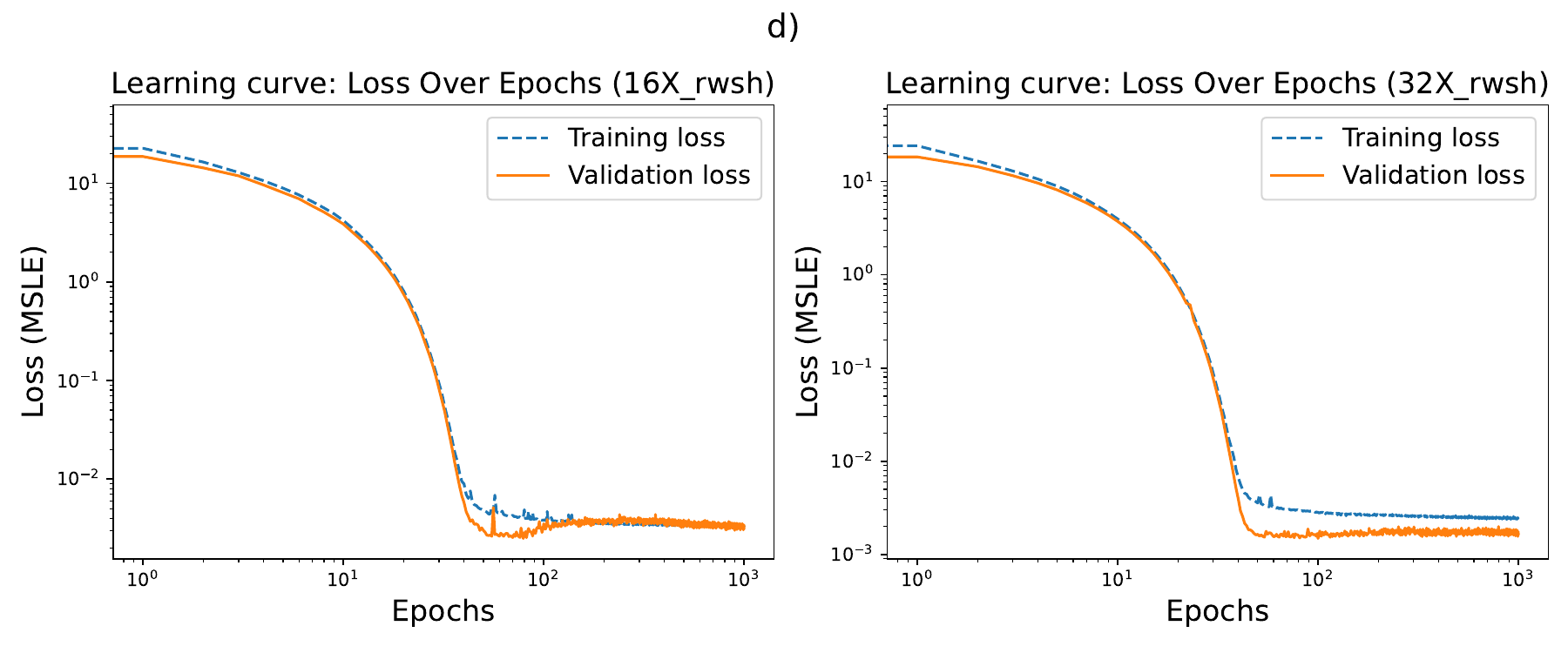}
\caption{Learning curves of our DNN-3 models. The initial dataset is divided into two parts: $80\%$ as training dataset and $20\%$ as test dataset. The training dataset is then divided anew into two parts: $80\%$ as the final training dataset and $20\%$ as the validation dataset.}
\label{learning_curves_results}
\end{figure*}

Furthermore, the \textit{TensorFlow} framework allows us to capture the history of the loss function  during training and make the learning curve of a \textit{DNN} model. We present these curves in Fig.~\ref{learning_curves_results}. The validation set provides insight into the stability of training by enabling direct comparison between the training loss and the validation loss. In our analysis, both the MSLE training loss and validation loss start from a few dozen and become very small at approximately 600 epochs, as shown in a) and c) of FIG. \ref{learning_curves_results}, with semi-log scaling. The log-log scaling in b) and d) of the same figure allows us to observe this behavior in more detail: there is a reduction in losses by four orders of magnitude. Additionally, the validation loss seems to be constantly (or mostly) below the training loss, implying that the \textit{DNN-3} models would perform better (or at least the same) on unseen data, rather than on the training dataset. 

For the neutron stars, the training loss ultimately falls behind the validation loss and the validation loss becomes continuously bigger after 400 epochs, indicating that the model begins to lose its accuracy (see b) in Fig.~\ref{learning_curves_results}). In contrast, for quark stars, the validation loss is always smaller than the training loss (see d) in Fig.~\ref{learning_curves_results}), ensuring better performance on foreign quark stars data.

A study of the effectiveness of statistical learning procedures would be incomplete without a reference to fitting times. As discussed in subsection \ref{fine_tune}, for the machine learning algorithms we employed 5-fold cross-validation and grid search. All computations were executed on an \href{https://www.intel.com/content/www/us/en/products/sku/236849/intel-core-ultra-9-processor-185h-24m-cache-up-to-5-10-ghz/specifications.html}{Intel Ultra 9 185H} CPU, using 18 of the 22 available threads. The fitting times are presented in Table \ref{fit_times}. As expected, the faster algorithm is \textit{Decision Tree}, due to its simple structure. In contrast, the slower algorithm is \textit{Gradient Boosting}, exceeding the threshold of 1 hour in the most extreme case of quark stars data with 32 feature variables.

\begin{table}[htb]
    \centering
    \begin{tabular}{|p{1.5cm}|p{1.5cm}|p{1.5cm}|p{1.5cm}|p{1.5cm}|}
    \hline
    \hline  
    \textbf{Model} & \textbf{16X NS} & \textbf{32X NS} & \textbf{16X QS} & \textbf{32X QS} \\
    \hline
    \hline
    Decision Tree  & $\approx 20^{''}$ & $\approx 30^{''}$ & $\approx 70^{''}$ & $\approx 2^{'}$\\
    \hline
    Random Forest  & $\approx 3^{'}20^{''}$ & $\approx 6^{'}$ & $\approx 14^{'}$ & $\approx 25^{'}$\\
    \hline
    Gradient Boosting  & $\approx 14^{'}15^{''}$ & $\approx 17^{'}30^{''}$ & $\approx 47^{'}30^{''}$ & $\approx 61^{'}30^{''}$ \\
    \hline
    XGBoost  & $\approx 3^{'}$ & $\approx 5^{'}30^{''}$ & $\approx 5^{'}30^{''}$ & $\approx 10^{'}30^{''}$ \\
    \hline
    DNN-3  & $\approx 4^{'}40^{''}$ & $\approx 5^{'}$ & $\approx 12^{'}$ & $\approx 12^{'}30^{''}$ \\
    \hline
    \hline     
    \end{tabular}
    \caption{Total fitting times of all regression models. For the machine learning algorithms times refer to 5-fold cross-validation and grid search combined. Columns 2 and 3 represent fitting times on Neutron Stars data, with 24200 rows of training data ($80\%\times 30400$) and 16 or 32 feature variables, respectively. Columns 4 and 5 represent fitting times on Quark Star data, with 71300 rows of training data ($80\%\times 89100$) and with 16 or 32 feature variables.}
    \label{fit_times}
\end{table}

\begin{figure*}[h]
\centering
\includegraphics[width=0.475\textwidth]{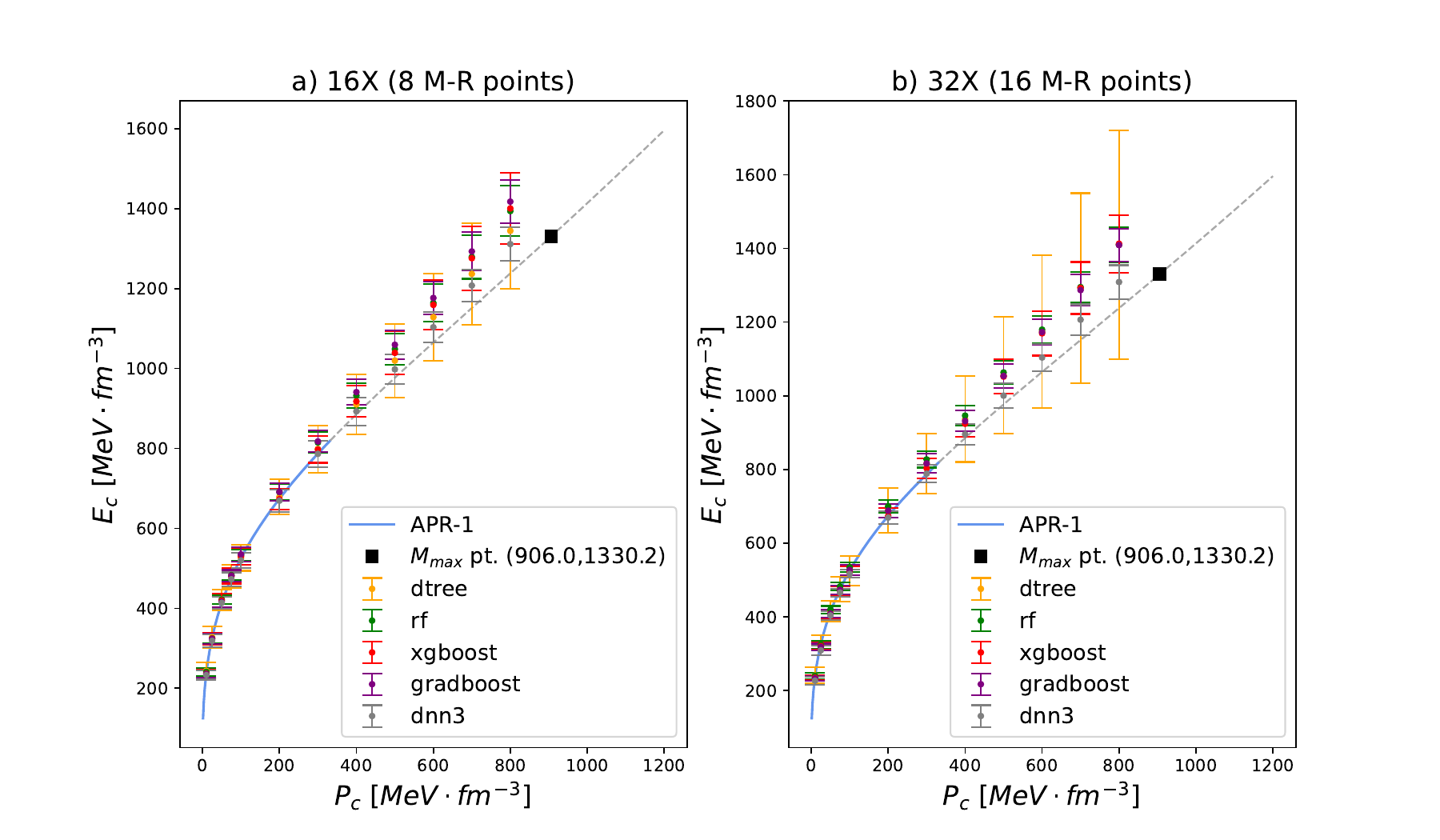}
\includegraphics[width=0.475\textwidth]{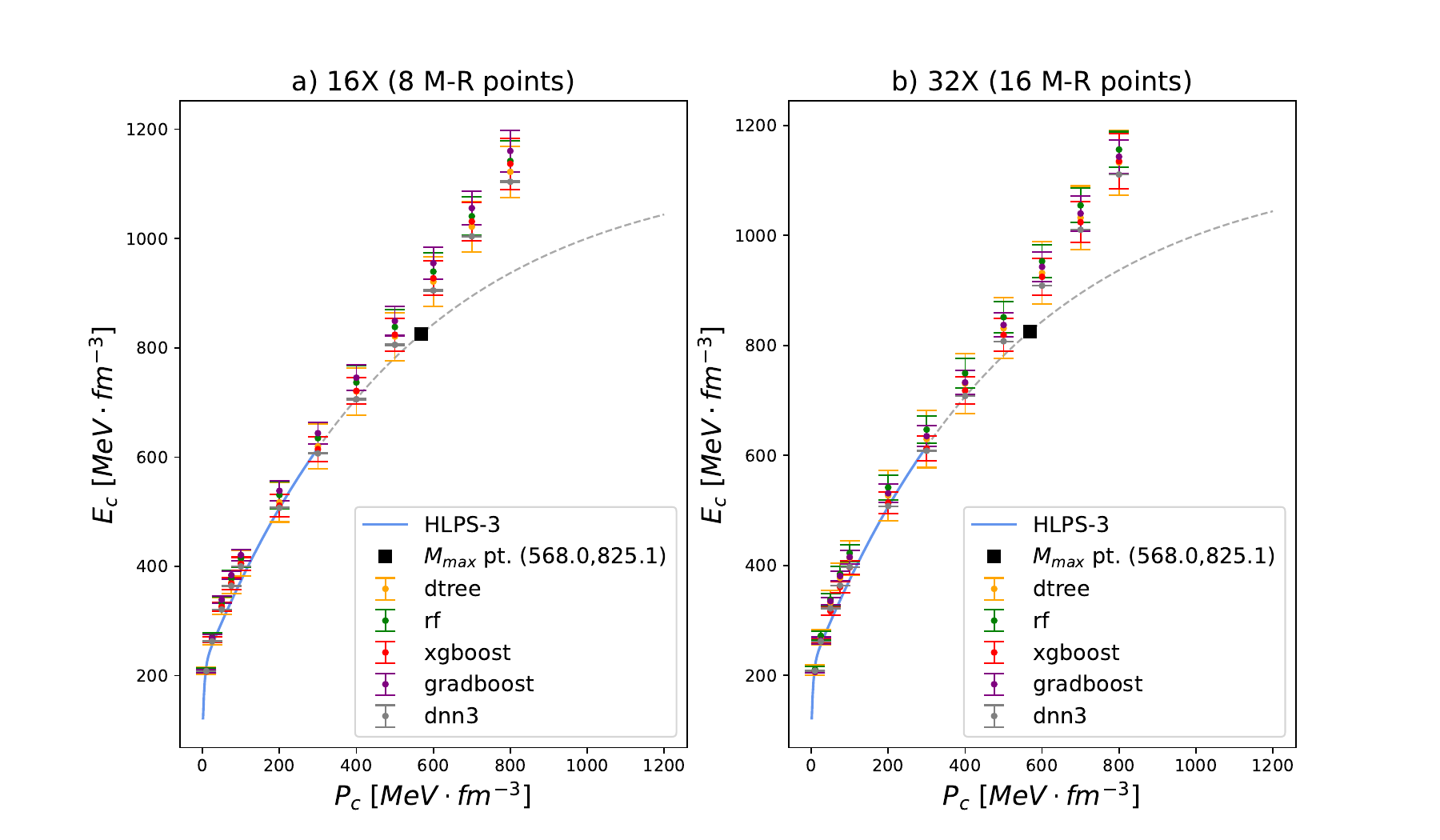}
\includegraphics[width=0.475\textwidth]{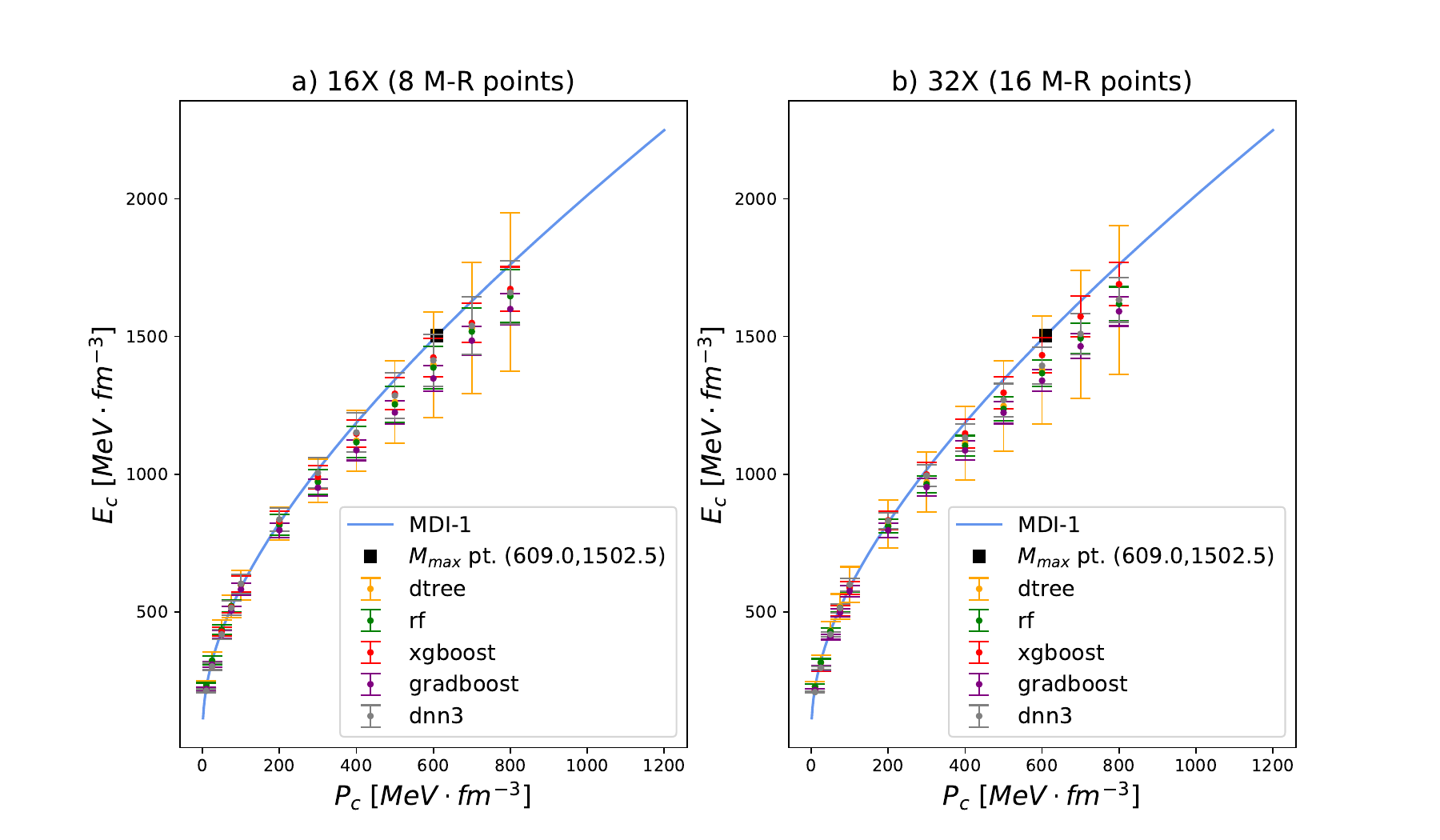}
\includegraphics[width=0.475\textwidth]{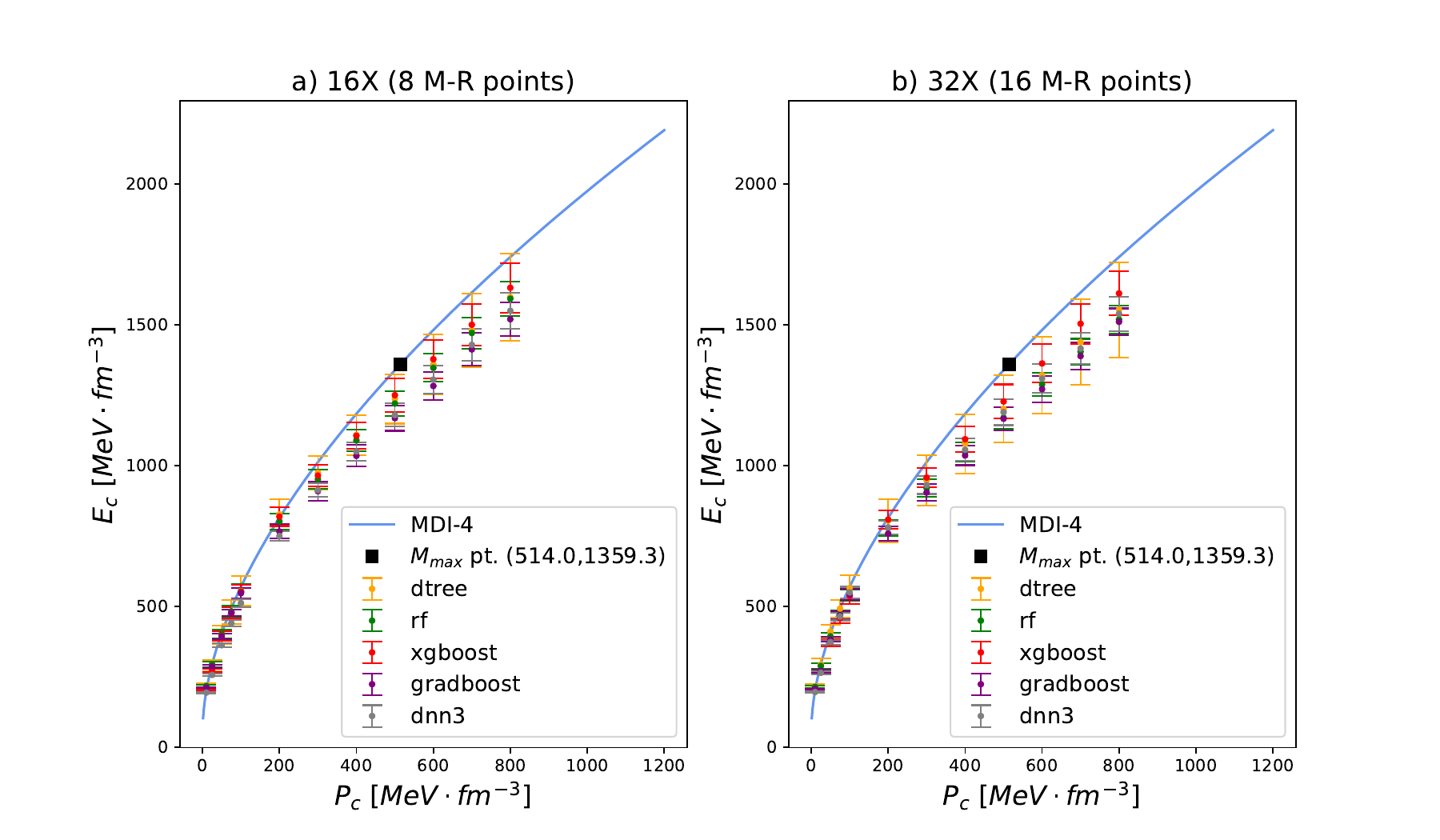}
\includegraphics[width=0.475\textwidth]{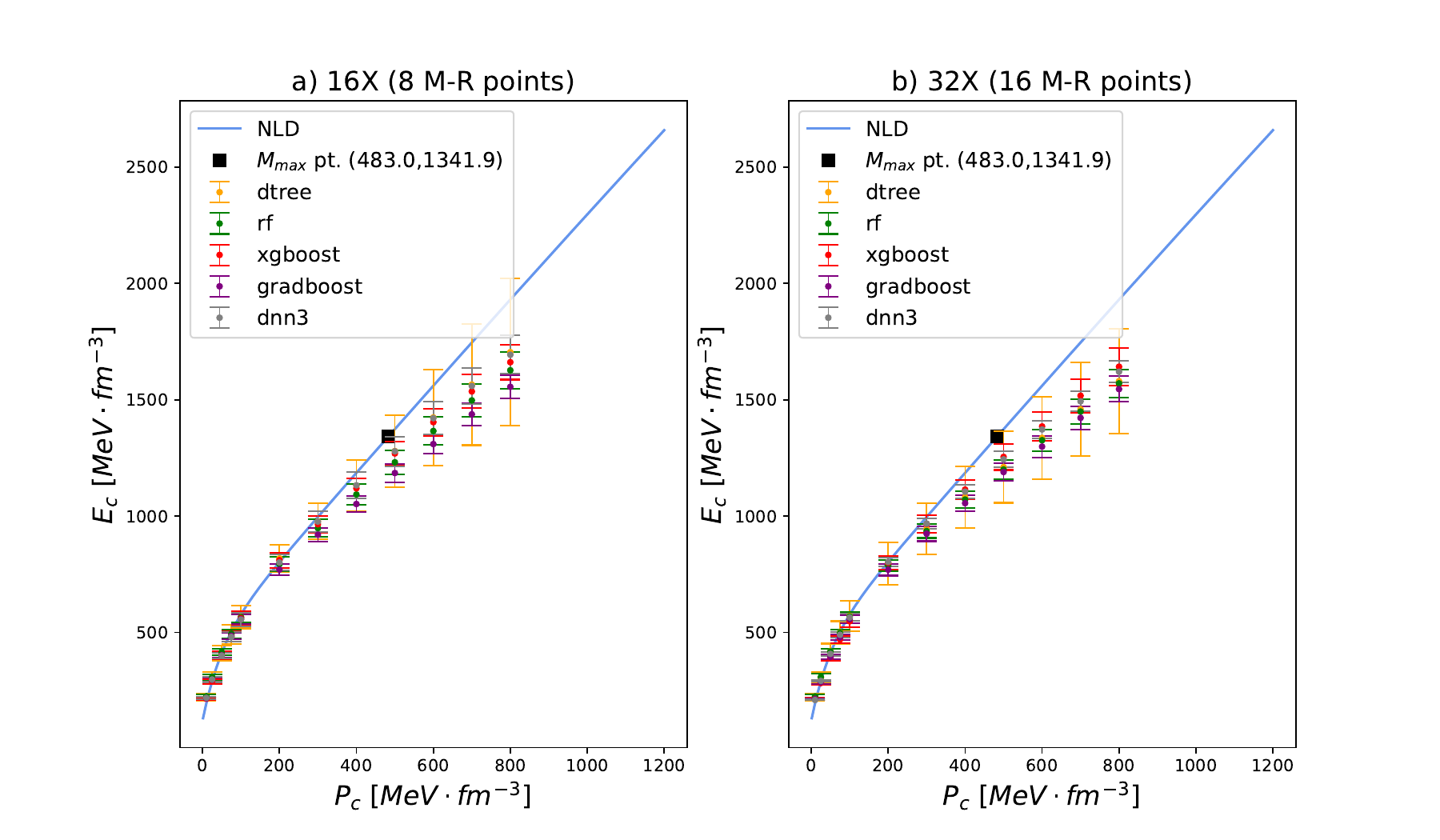}
\includegraphics[width=0.475\textwidth]{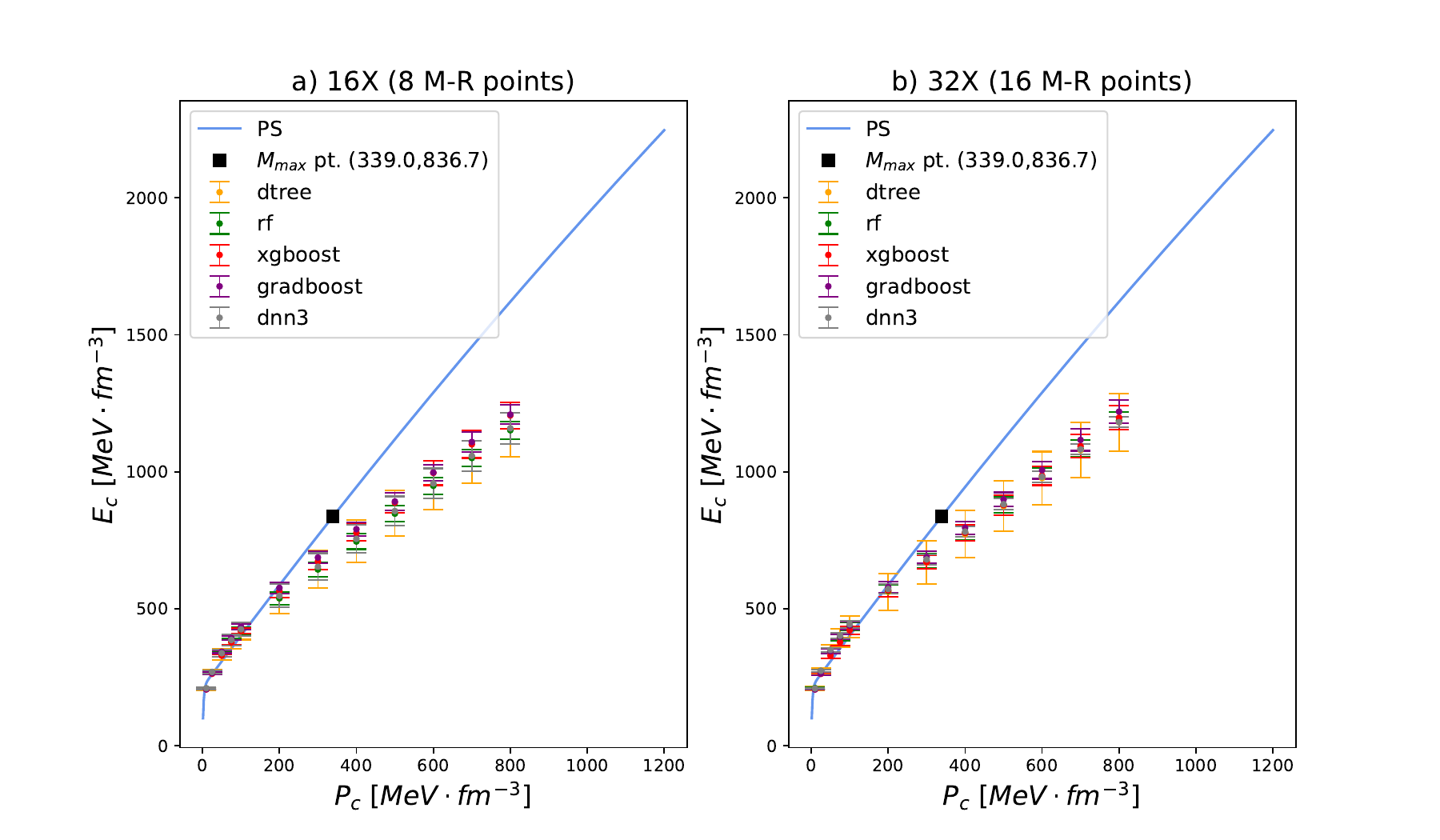}
\includegraphics[width=0.475\textwidth]{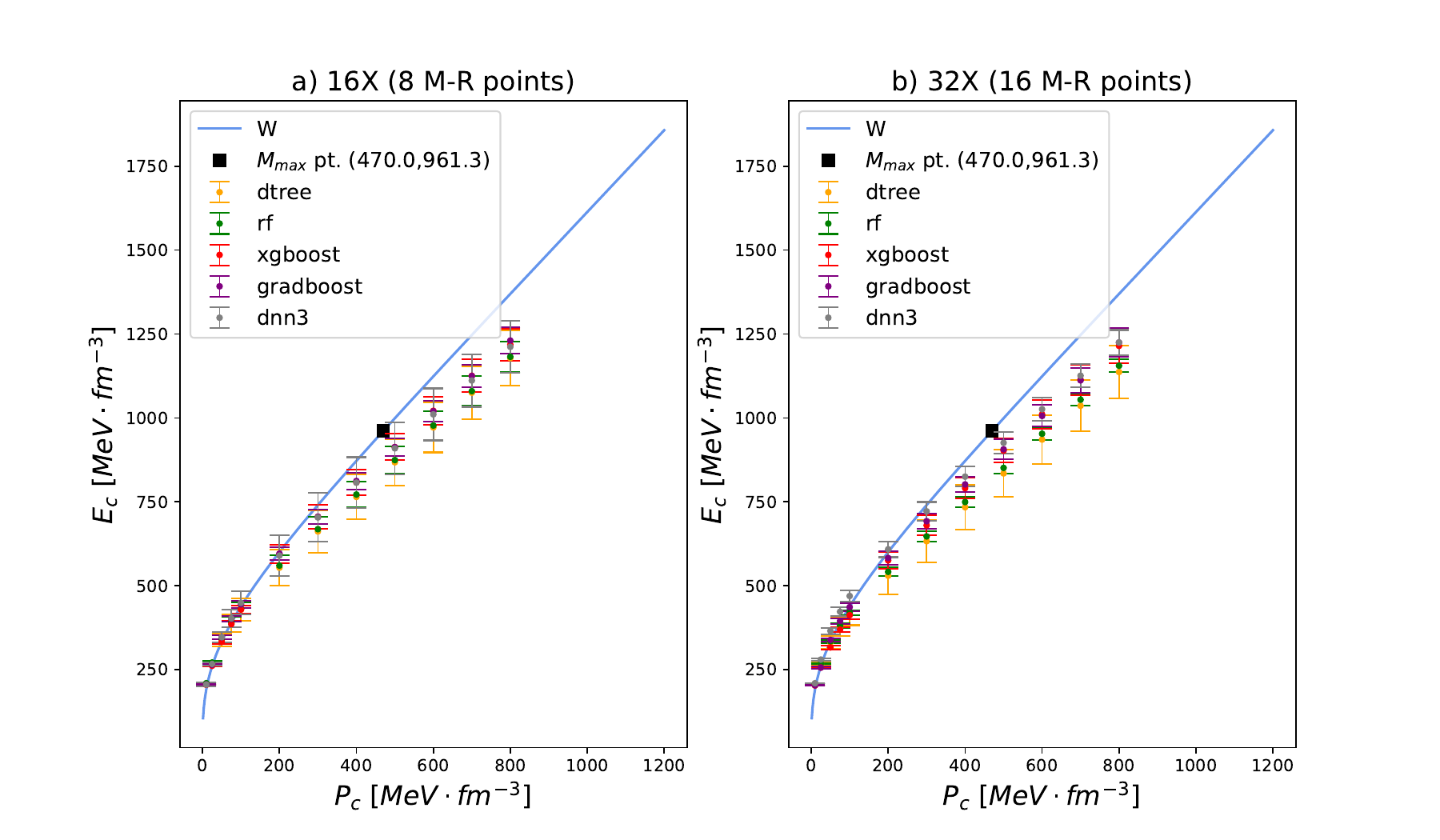}
\includegraphics[width=0.475\textwidth]{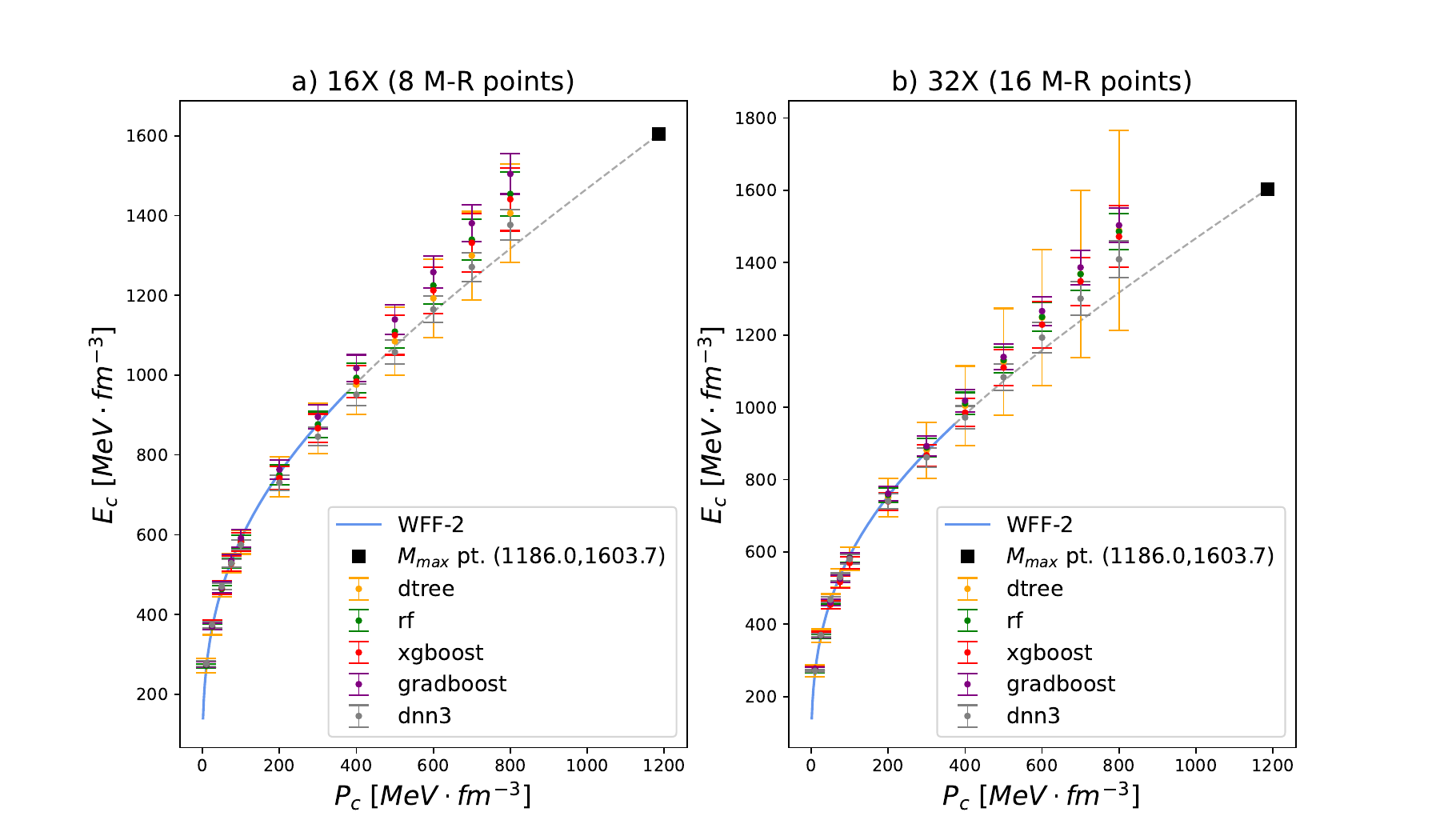}
\caption{GROUP A of reconstruction results for hadronic EoSs. The predictions of four machine learning models (\textit{Decision Tree, Random Forest, Gradient Boosting, XGBoost}) and our Artificial Neural Network model (see \ref{fine_tune}) are shown with different colors. For each model, the mean values and the error bars (standard deviation) of its energy density predictions are depicted. The actual EoS is, also included as solid line, for direct comparison and accuracy assessment. The black square denotes the  maximum mass point. }
  \label{group_A_predict_NS}
\end{figure*}

\begin{figure*}[h]
  \centering
 \includegraphics[width=0.475\textwidth]{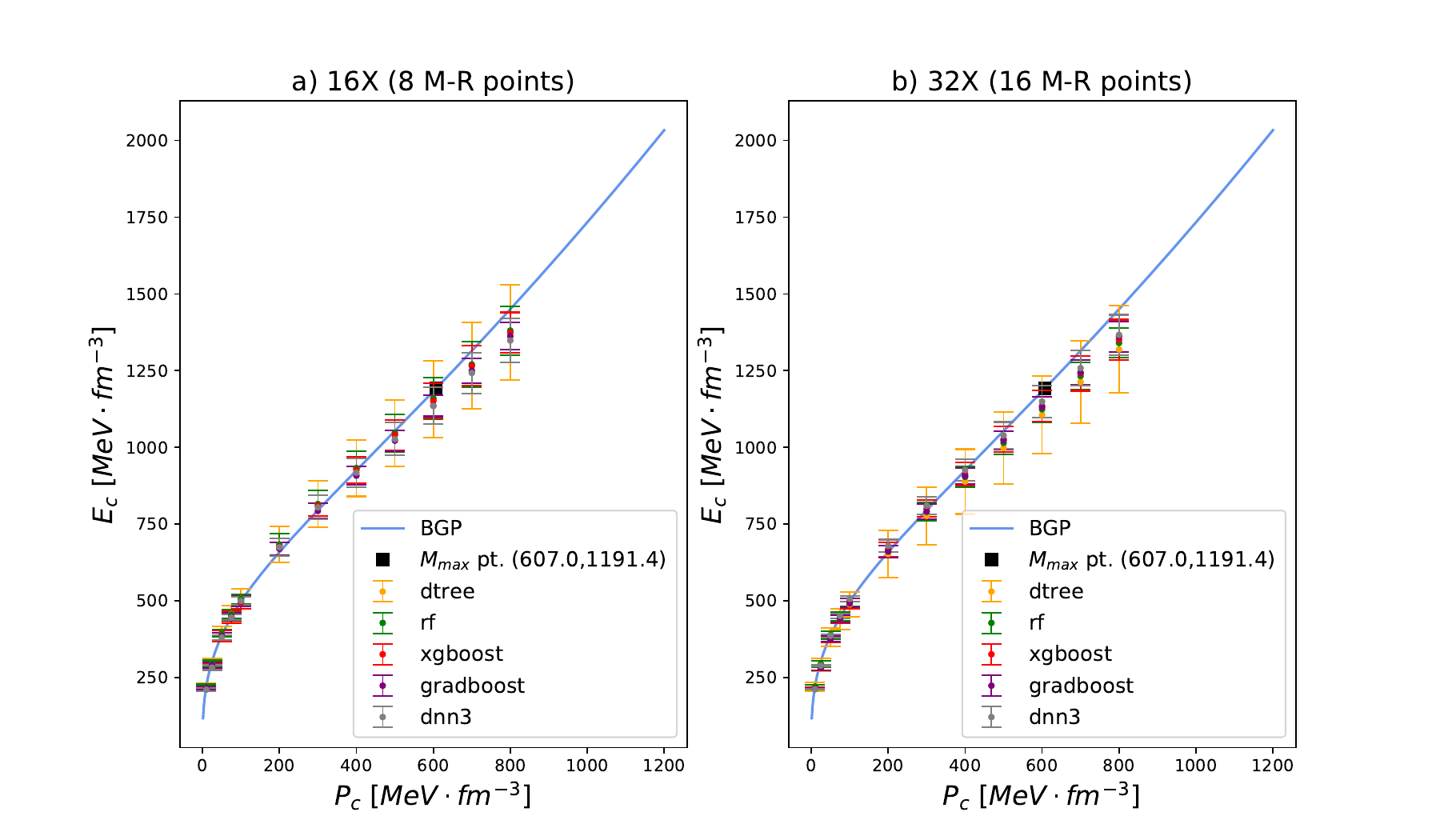}
  \includegraphics[width=0.475\textwidth]{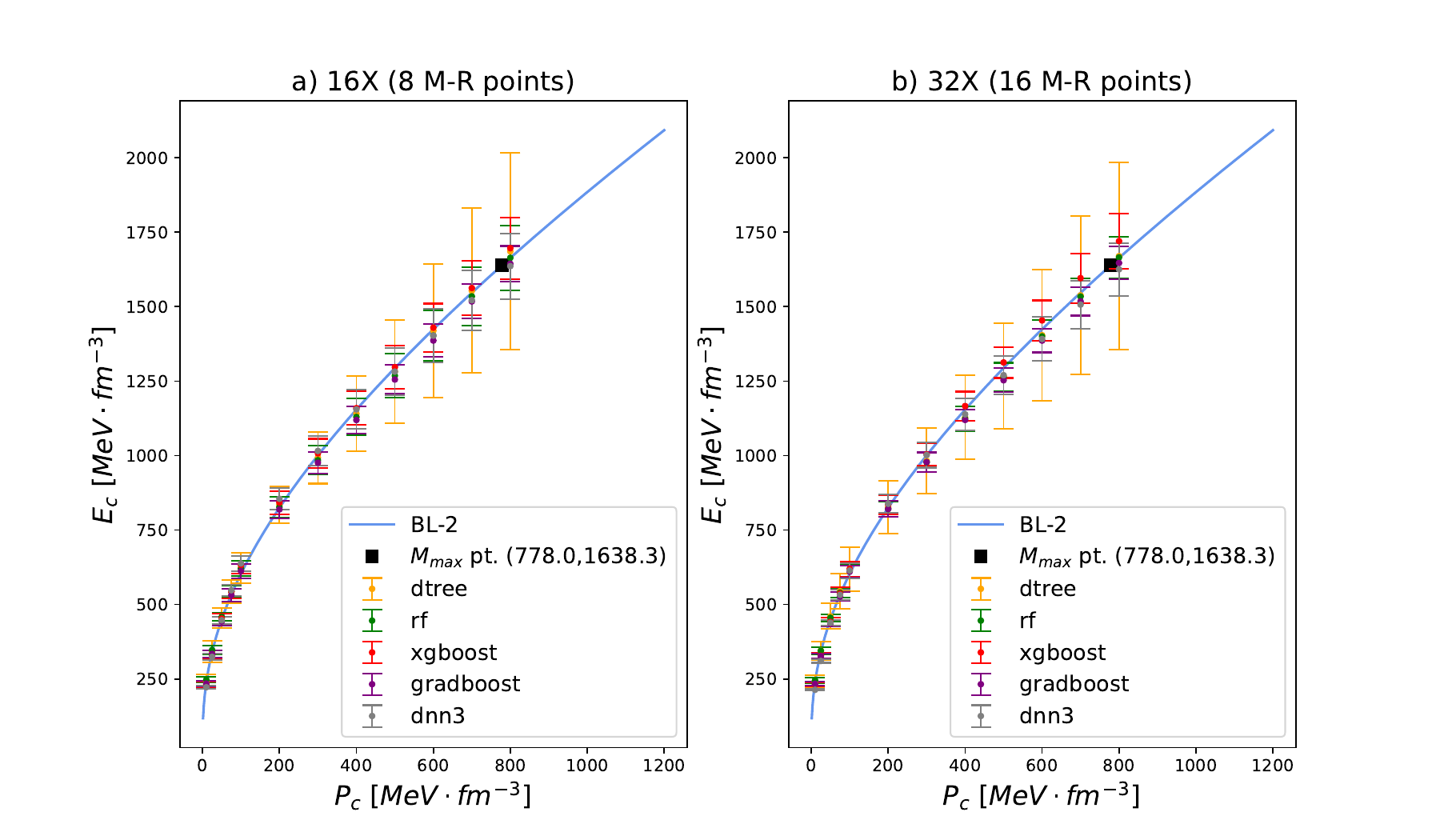}
  \includegraphics[width=0.475\textwidth]{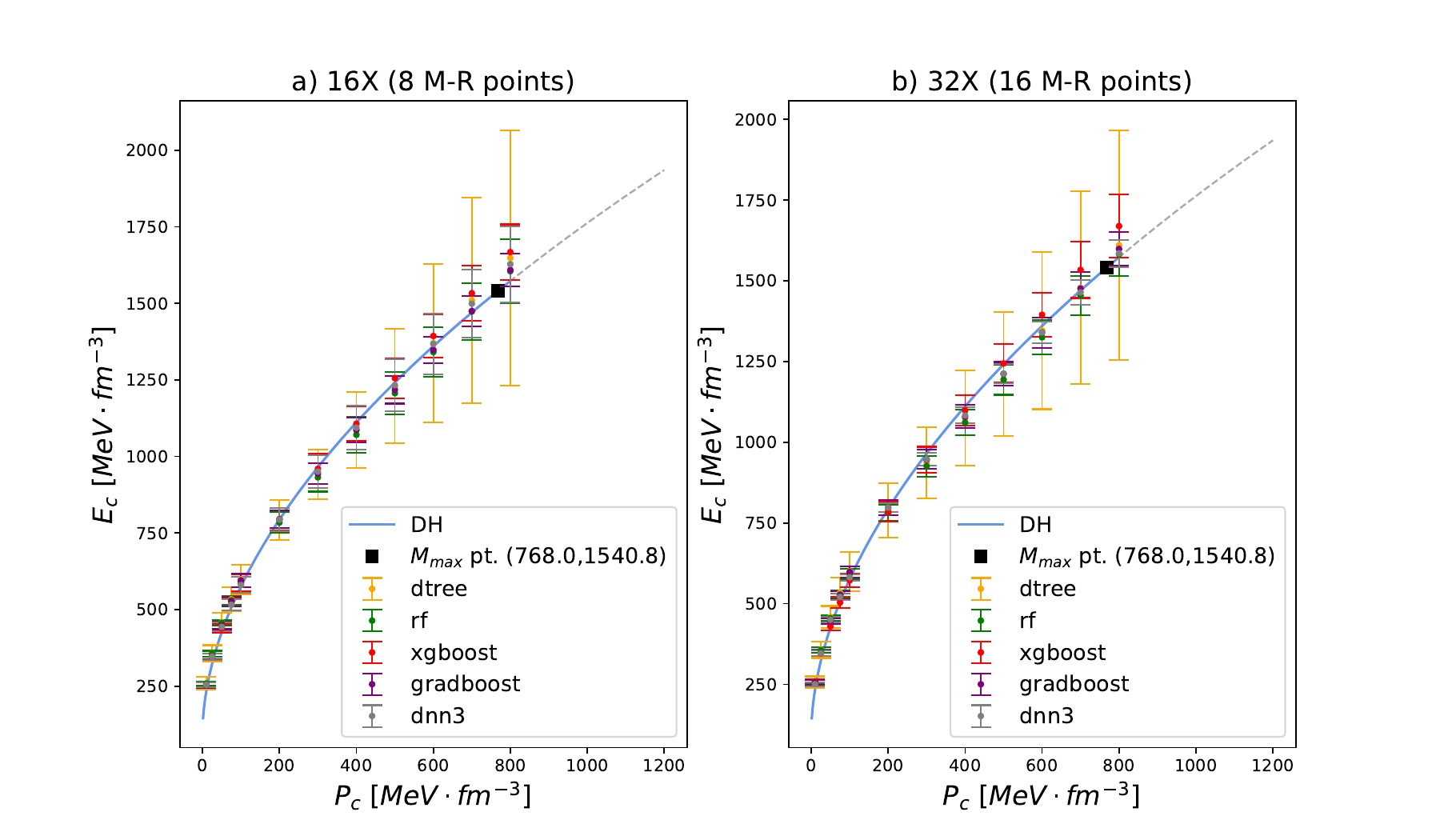}
  \includegraphics[width=0.475\textwidth]{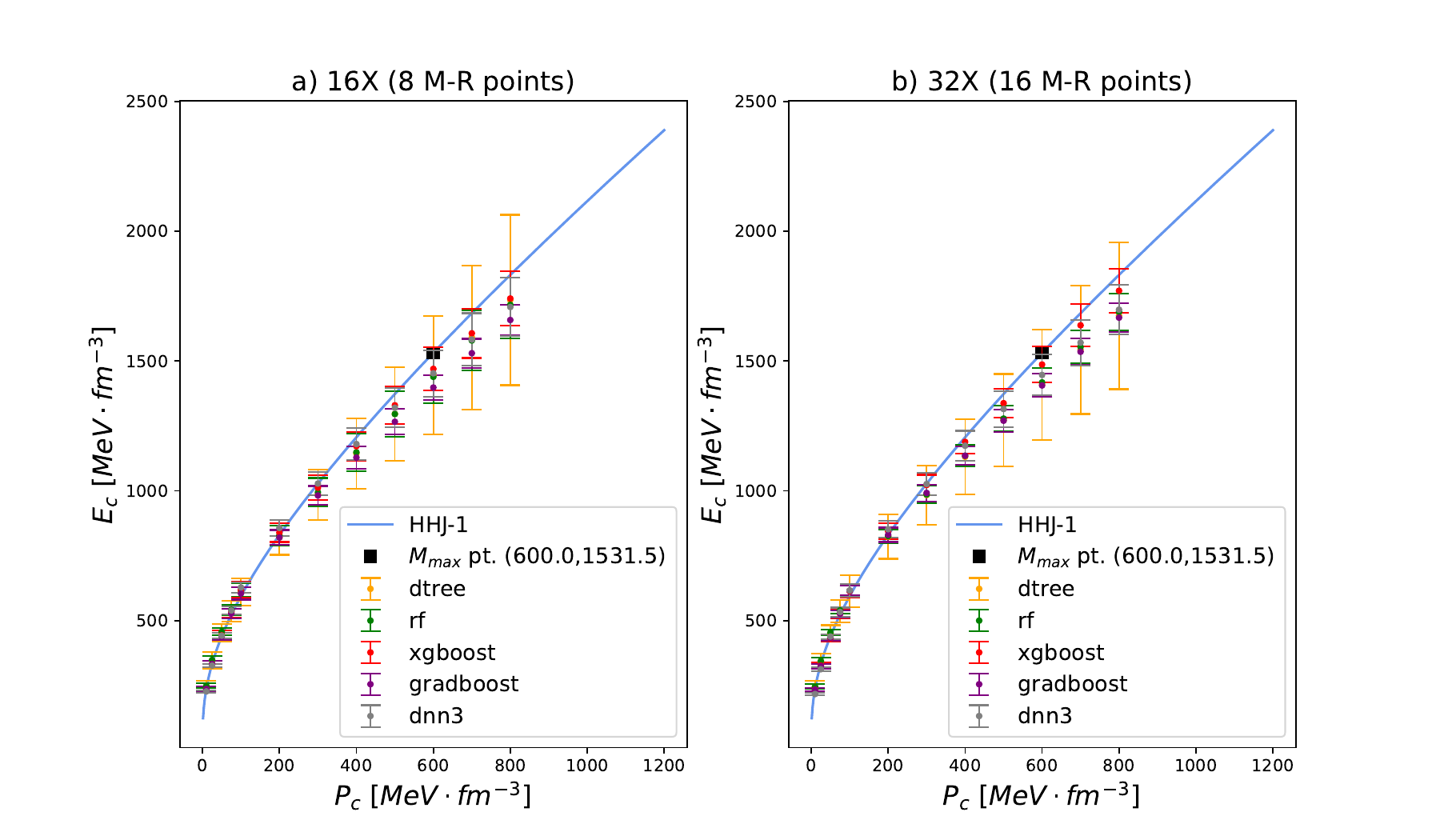}
  \includegraphics[width=0.475\textwidth]{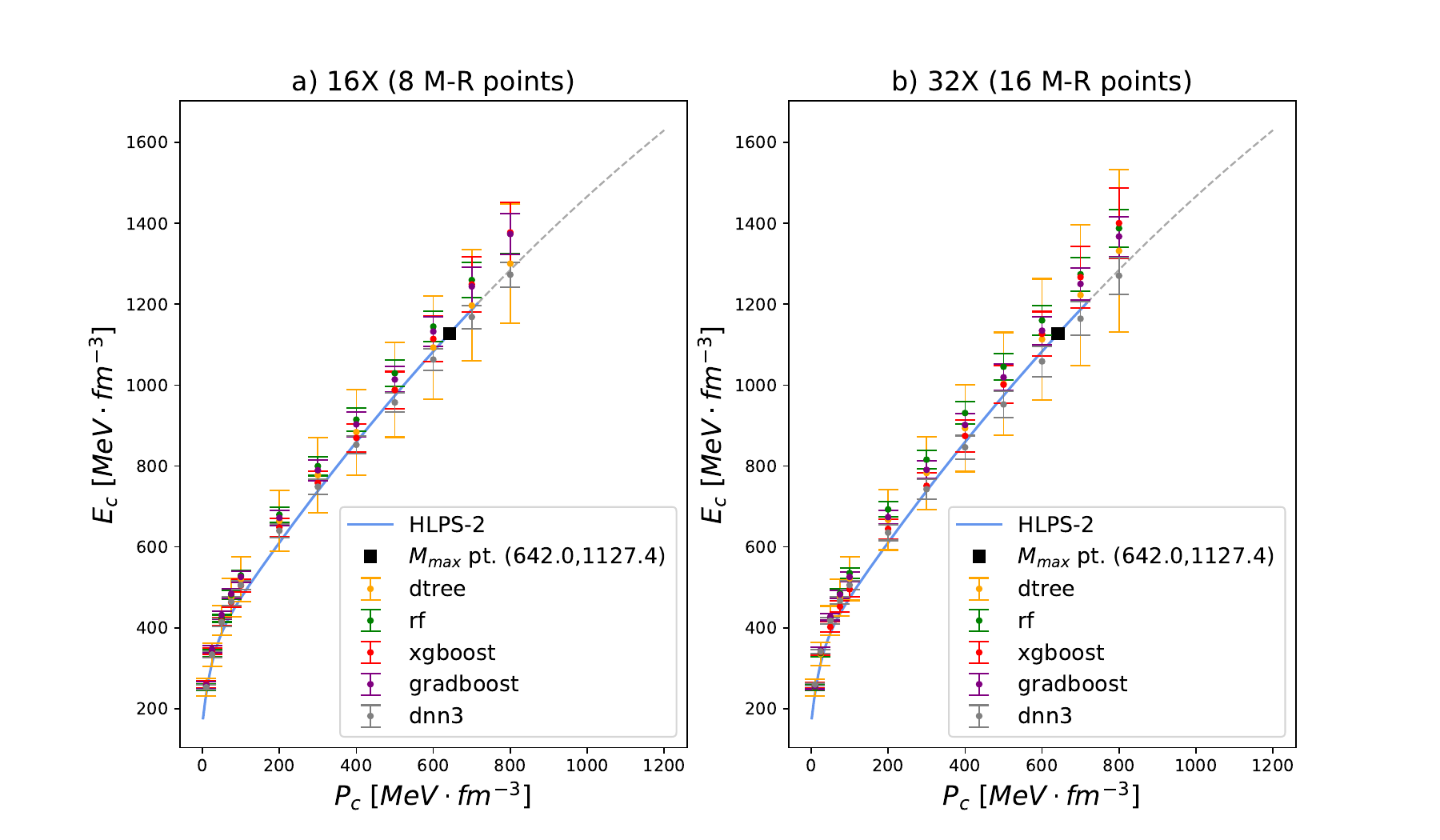}
  \includegraphics[width=0.475\textwidth]{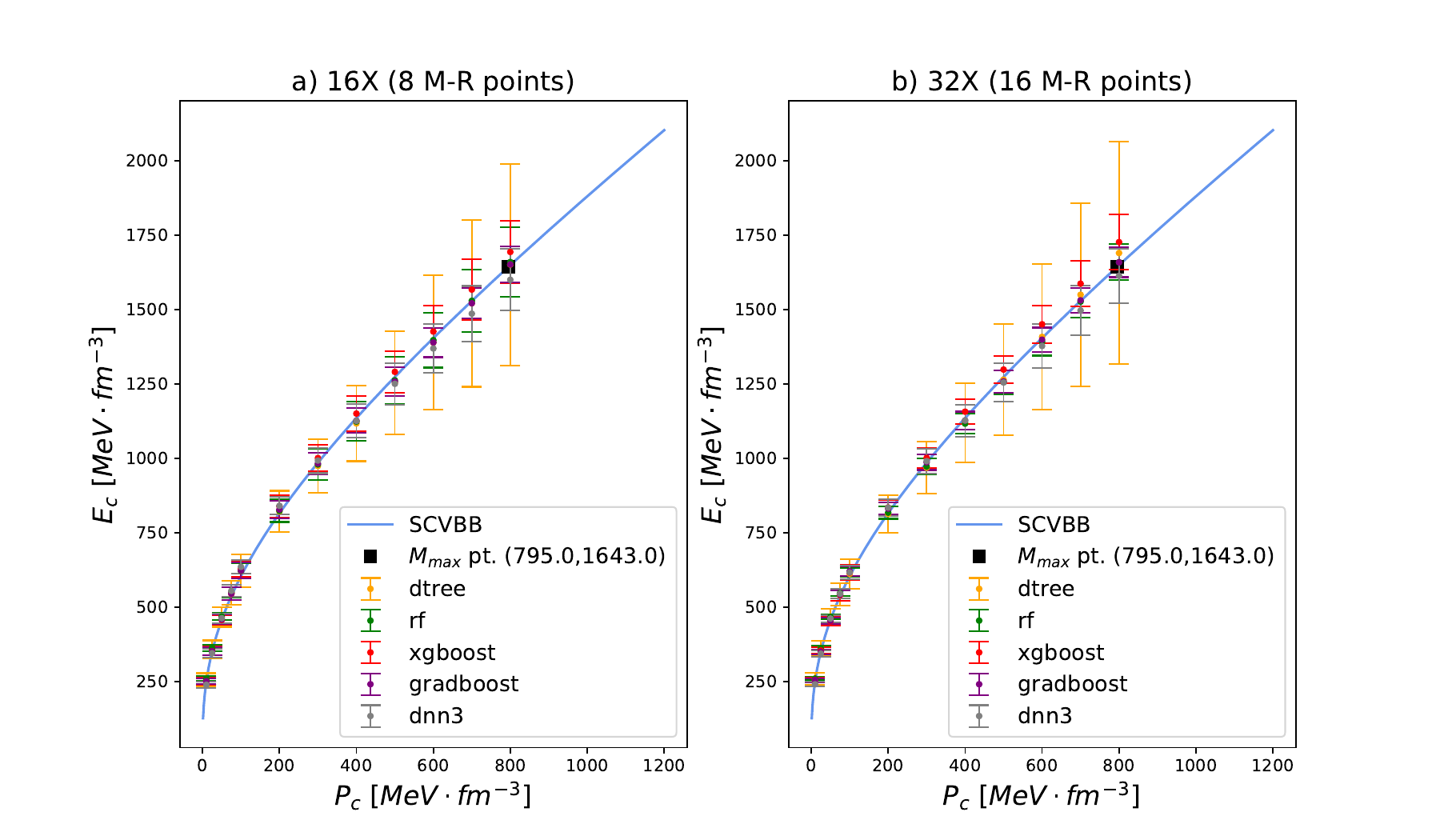}
  \includegraphics[width=0.475\textwidth]{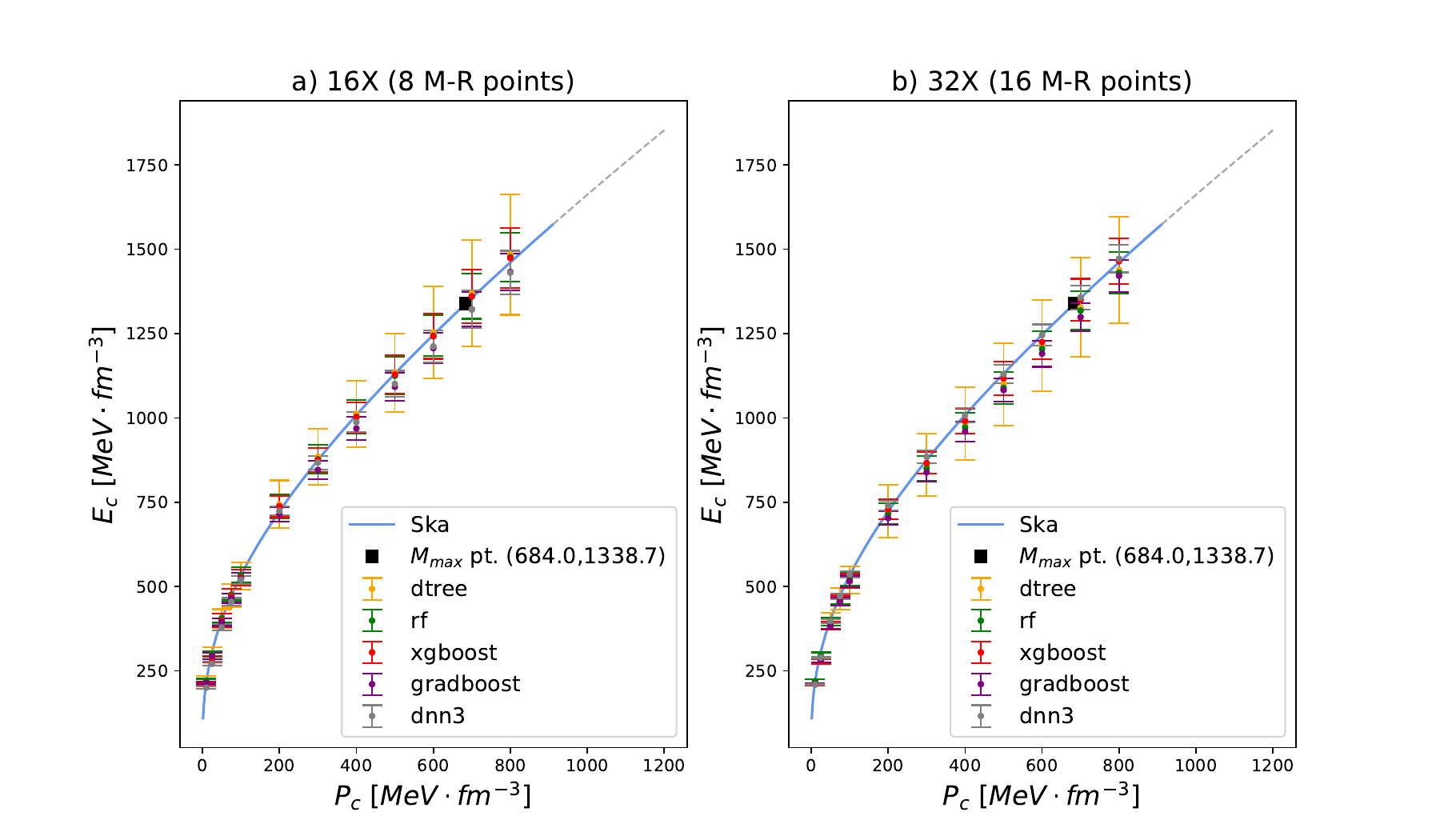}
  \includegraphics[width=0.475\textwidth]{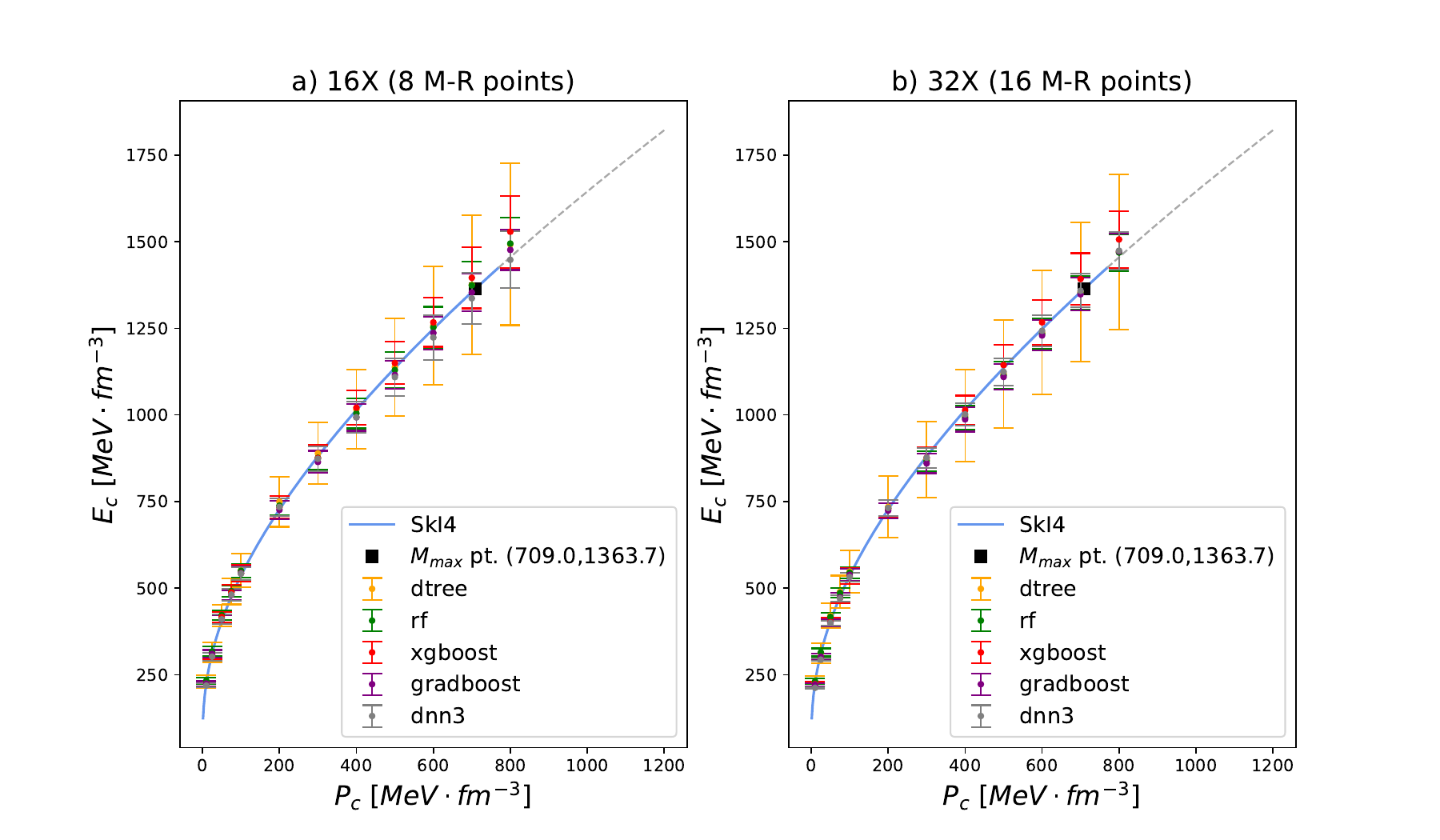}
\caption{GROUP B of reconstruction results for hadronic EoSs. Same as FIG. \ref{group_A_predict_NS}, but for different hadronic EoSs.}
  \label{group_B_predict_NS}
\end{figure*}

\begin{figure*}[h]
  \centering
  \includegraphics[width=0.475\textwidth]{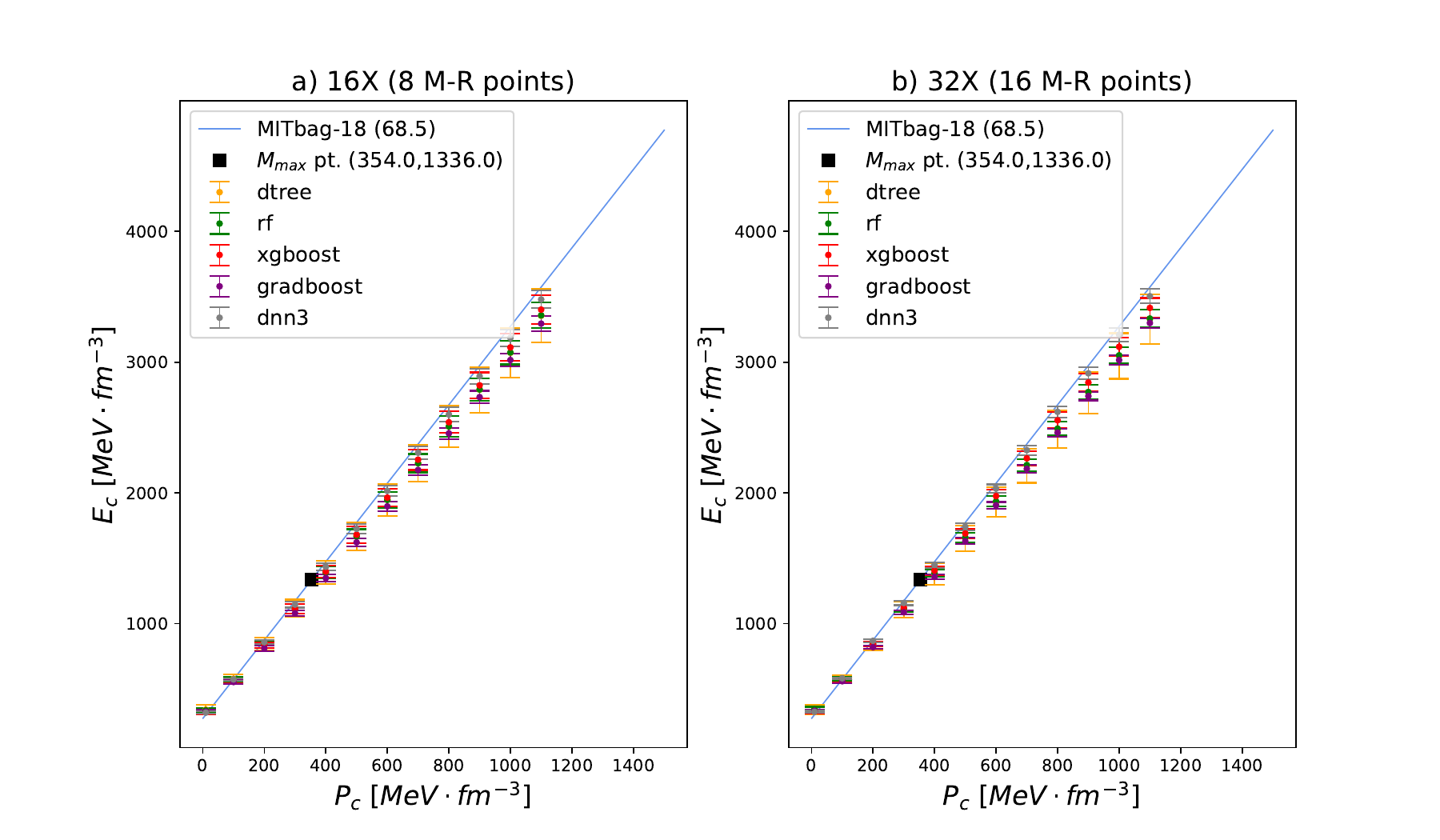}
  \includegraphics[width=0.475\textwidth]{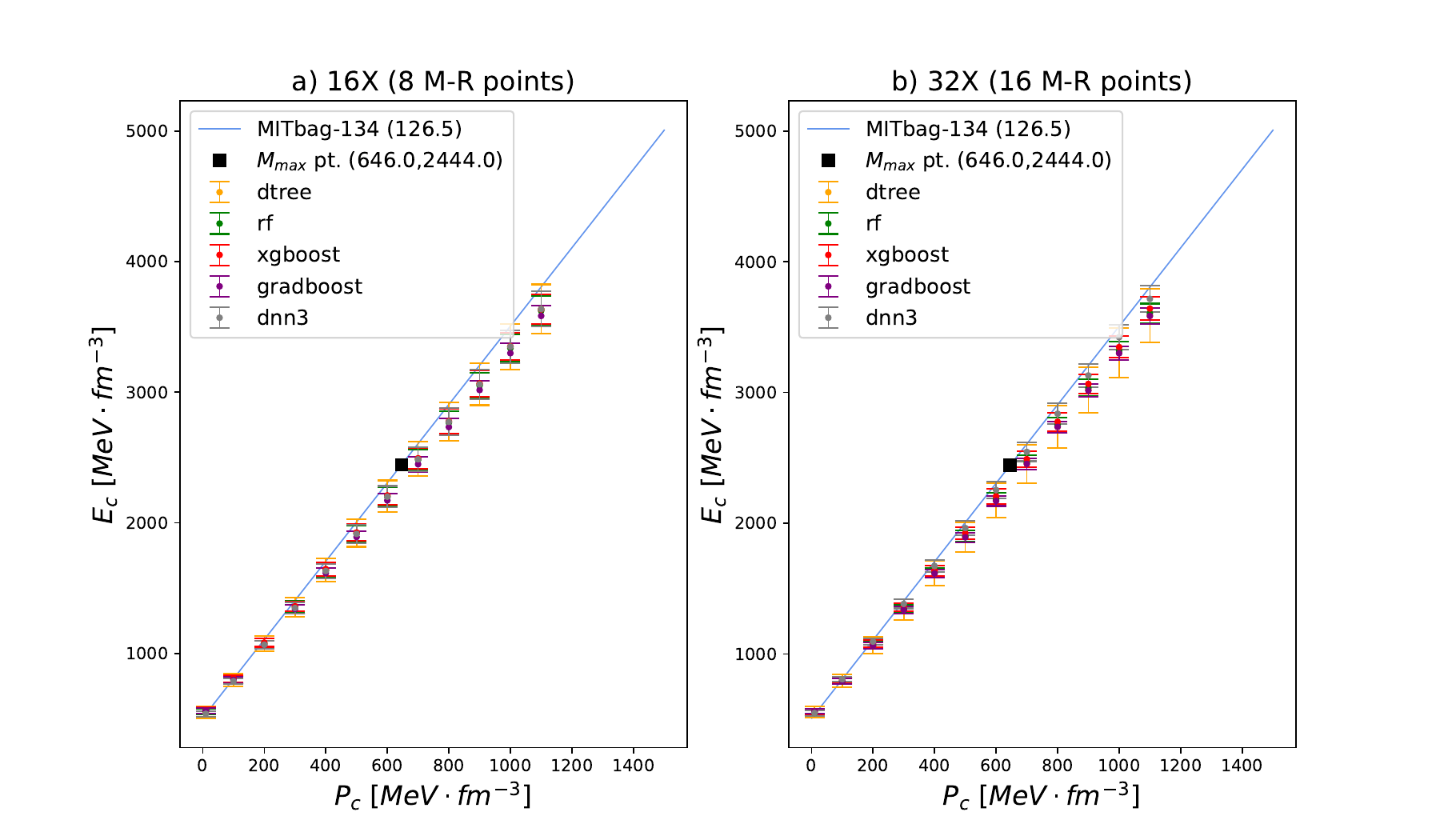}
  \includegraphics[width=0.475\textwidth]{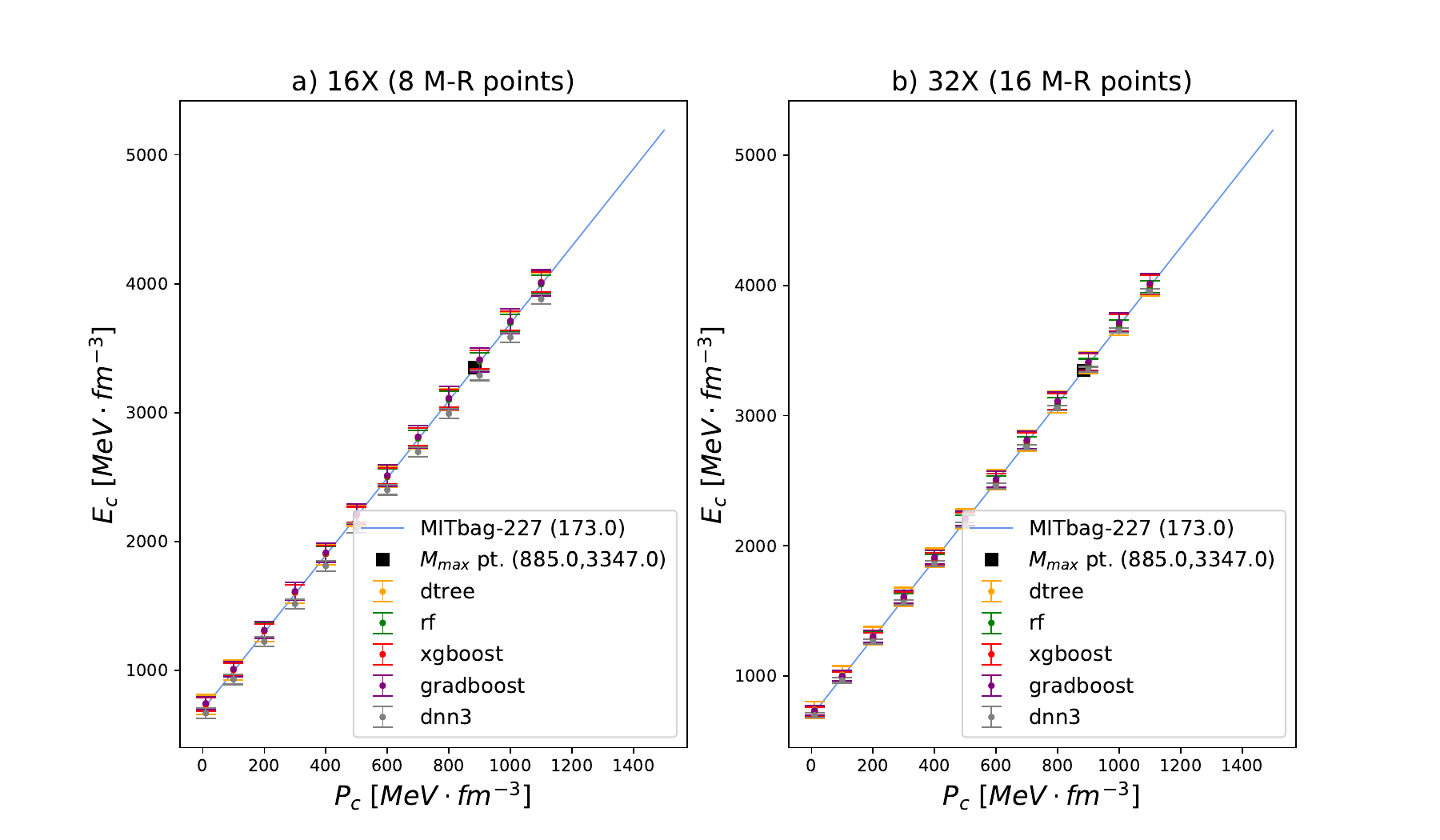}
  \includegraphics[width=0.475\textwidth]{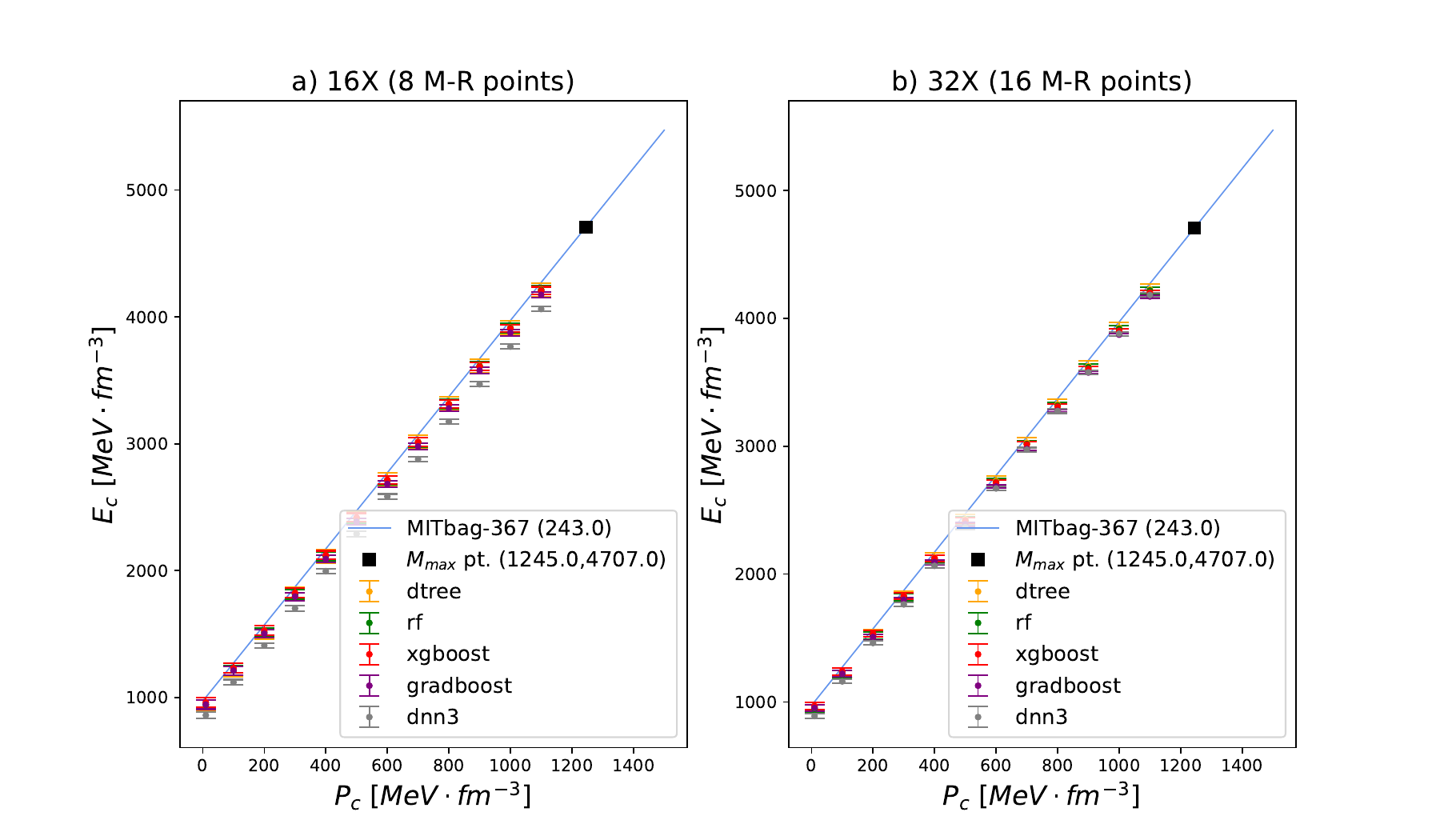}
\caption{Reconstruction results of MIT-bag EoSs. Same as FIG. \ref{group_A_predict_NS}, but for quark-matter EoSs derived from the MIT-bag model.}
  \label{MITbag_predict}
\end{figure*}

\begin{figure*}[h]
  \centering
  \includegraphics[width=0.475\textwidth]{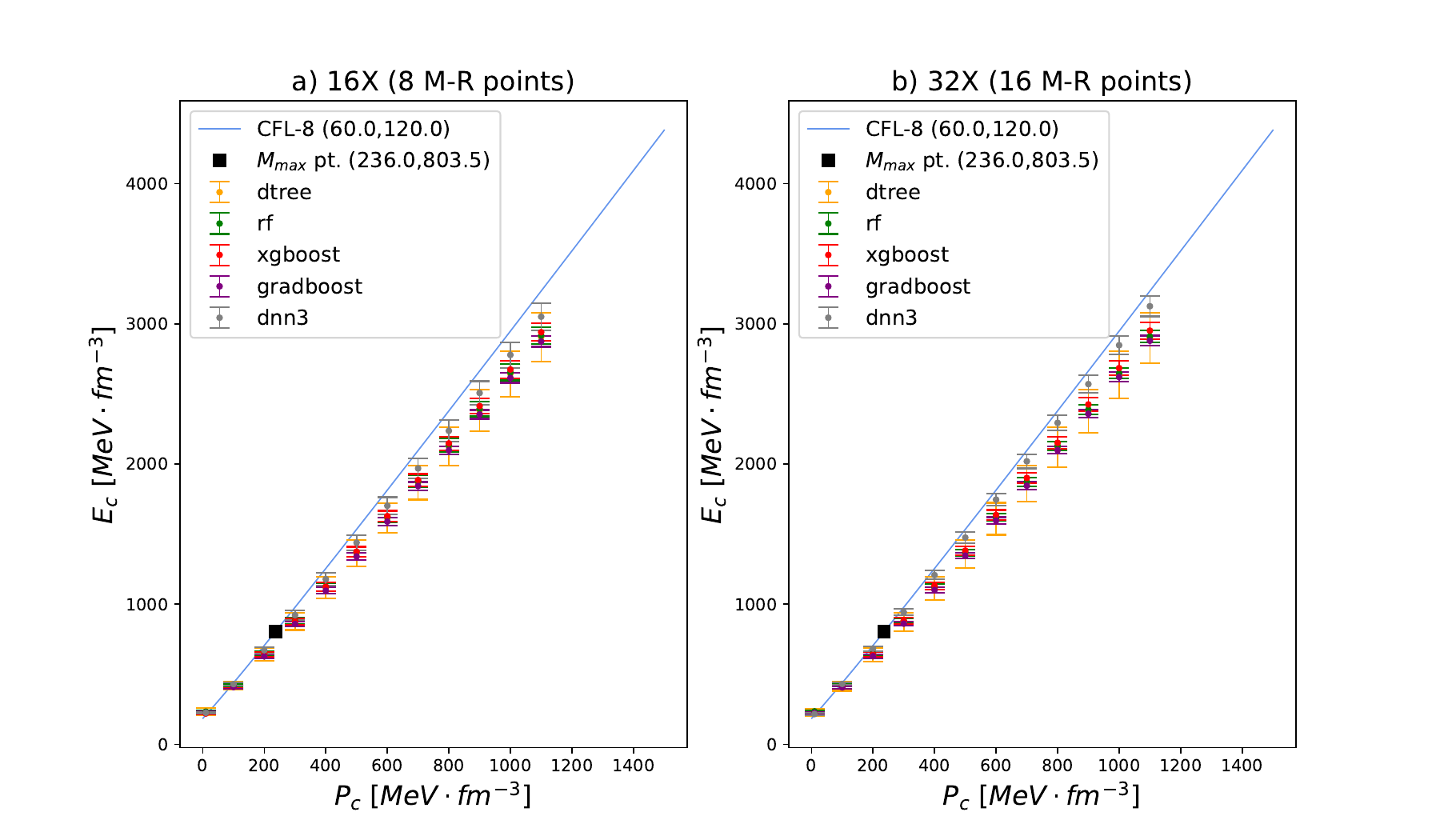}
  \includegraphics[width=0.475\textwidth]{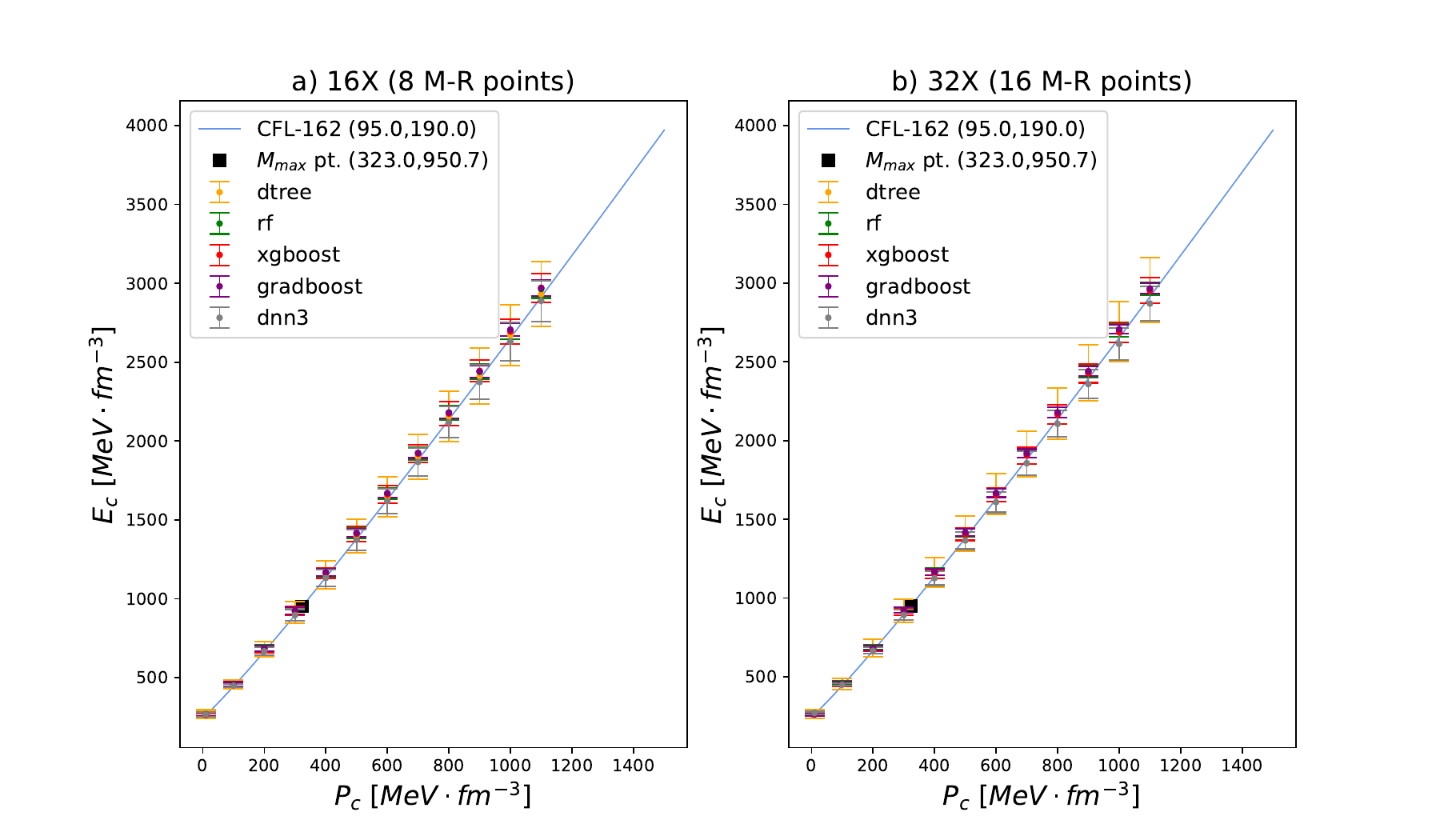}
  \includegraphics[width=0.475\textwidth]{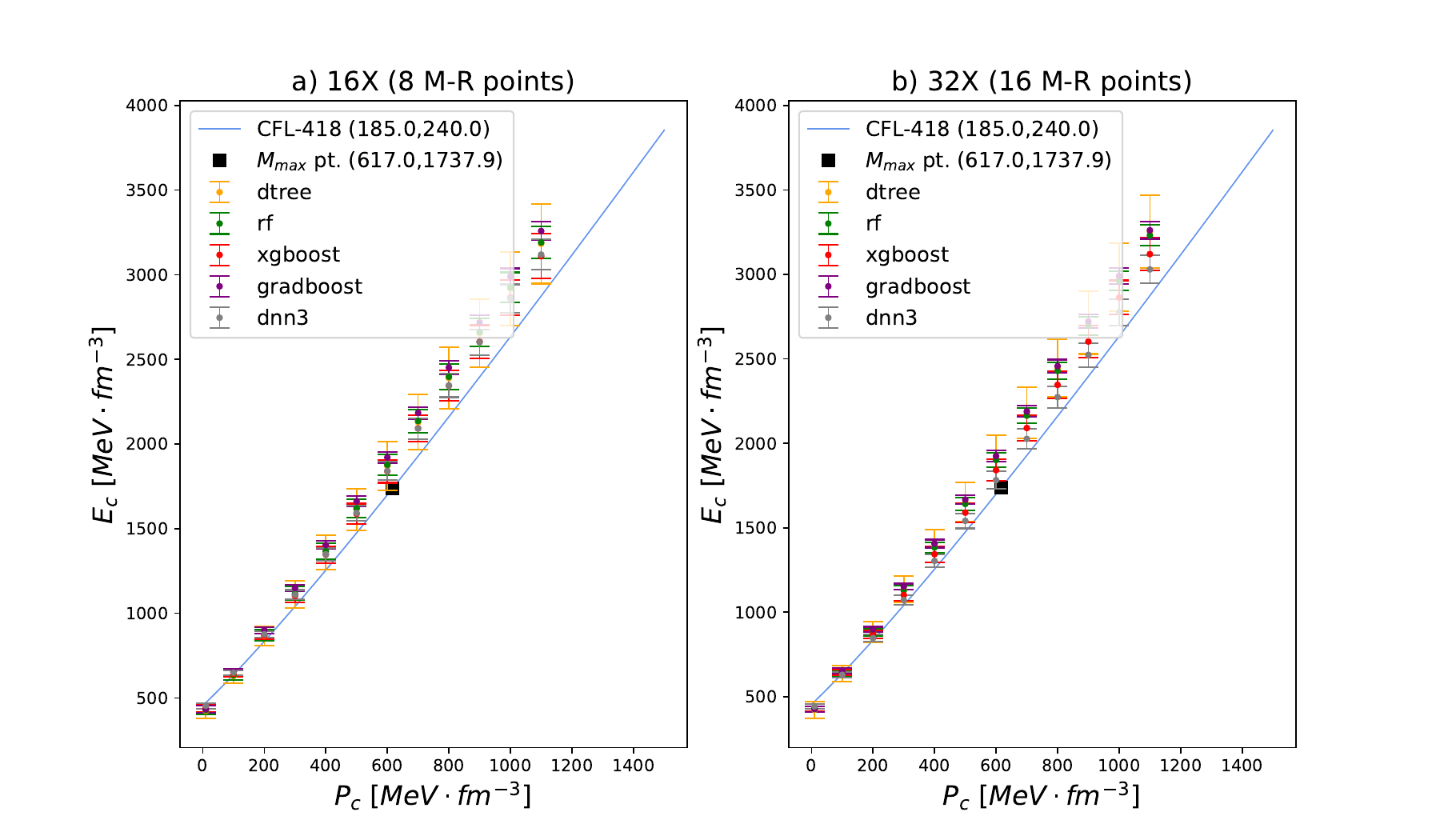}
  \includegraphics[width=0.475\textwidth]{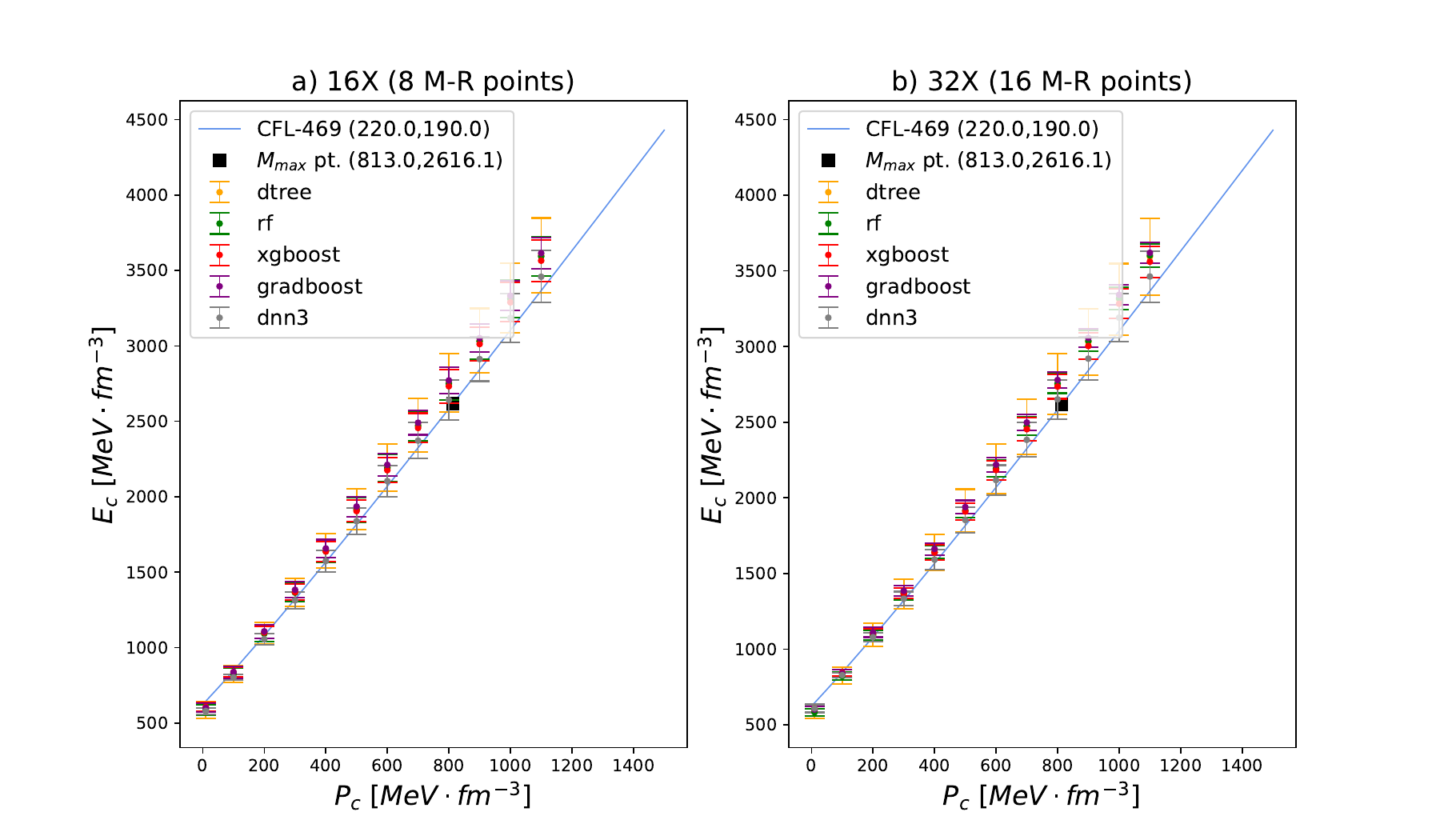}
\caption{Reconstruction results of CFL EoSs. Same as FIG. \ref{group_A_predict_NS}, but for quark-matter EoSs derived from the CFL model.}
  \label{CFL_predict}
\end{figure*}

We now proceed to outline the methodology employed to generate the plots presented in Figs..~ \ref{group_A_predict_NS}, \ref{group_B_predict_NS}, \ref{MITbag_predict} and \ref{CFL_predict}. To reconstruct each EOS, we obtained 100 random observations of $M-R$ points, from its respective $M-R$ curve. We shuffled them differently for each observation, creating samples of data similar to those used to fit the regression models. Then, we computed the energy density predictions at the selected pressure values (see subsection \ref{data_prep}), feeding each $M-R$ observation into our regression models. Thus, for each model, we obtained 100 predictions of the energy density per pressure value. We calculated the mean value and standard deviation of the 100 predictions and constructed error bars in order to assess the accuracy and variance of the respective model in reconstructing the EoS. In Figs.~\ref{group_A_predict_NS}, \ref{group_B_predict_NS}, \ref{MITbag_predict} and \ref{CFL_predict}, the error bars of the predictions from each regression model are depicted, along with the respective original EoS. The circle in the center of the error bars corresponds to the mean prediction, while the total length of the error bars is twice the standard deviation of the predictions.

Our first observation is similar to the overall training results in Figs.~\ref{metrics_results_NS} and \ref{metrics_results_QS}. Poor performance of the \textit{Decision Tree} models compared to all others. \textit{Decision Tree} models exhibit low accuracy, since the mean prediction deviates significantly from the original EoS and they also have high variance. In contrast, \textit{DNN-3} models have the minimum prediction variance among all. However, the \textit{DNN-3} model is not always the best algorithm for reconstruction: for a certain EoS, its mean predictions might deviate more from the original EoS than those of some machine learning models. As for the \textit{Random Forest}, \textit{Gradient Boosting} and \textit{XGBoost} models, these perform the same on average, as we denoted in figures. \ref{metrics_results_NS} and \ref{metrics_results_QS}, as well.

Furthermore, in graphs labeled a) in Figs.~\ref{group_A_predict_NS}-\ref{CFL_predict} the reconstruction is made using 8 points from the $M-R$ curves, while in graphs labeled b), the reconstruction uses 16 points from the $M-R$ curves. As seems in most cases, we obtain similar results, regardless of the number of points. In practice, the 16 M-R points offer little improvement. 

Moving on, we notice the excellent reconstruction of every EoS at lower pressure ($<100$ $\rm MeV\ fm^{-3}$). Furthermore, we denote the gradual but evident increase in variance as we move into high pressure for all algorithms. This confirms our discussion in section \ref{EOS_theory}. The equation of state is well-defined in low mass densities (lower than nuclear saturation density) and their formula becomes unknown as we reach extremely high mass densities, due to the uncertainty in the composition of matter at these densities. Nevertheless, we obtain reliable results till the maximum mass point (black square in figures): the mean predictions from all regression models are very close or coincide with the original EoS until that point. In other words, since the maximum mass point marks the transition between stable and unstable compact star configurations, we can claim that we reconstruct the stable part of the EoS with good accuracy.

In more specific results for neutron stars, the violation of causality might be a factor that affects the performance of the regression models. It should be emphasized that all algorithms are trained exclusively on datasets that satisfy the causality requirements. Hence, our models start to exhibit greater variance when entered regions of causality violation and the maximum mass point lies inside this region, as shown for the reconstruction of the EoSs \textbf{APR-1}, \textbf{HLPS-3} \textbf{WFF-2} (Fig.~ \ref{group_A_predict_NS}). On the other hand, the good reconstruction of many hadronic EoSs, shows that the behavior of these EoSs can indeed be reproduced by multimodal parameterization, as shown in Fig.~\ref{group_B_predict_NS}. In other words, polytropic parameterization is an effective technique for capturing the details of an EoS at all pressures. Of course, there might be exceptions, like these for the EoSs \textbf{MDI-1}, \textbf{MDI-4}, \textbf{NLD}, \textbf{PS} and \textbf{W} (Fig.~\ref{group_A_predict_NS}). It is also worth mentioning that algorithms that perform great with some EoSs, might perform poorly with other.

Finally, we address the reconstruction of quark matter EoSs. This exhibits much smaller variances compared to the reconstruction of hadronic EoSs. Due to the simpler and better-defined form of quark star EoSs. The latter is also the reason we get good results even after the maximum mass points. That is, we can effectively predict the unstable part of the EoSs region. Moreover, we observe the evident superiority of the \textit{DNN-3} models: these models exhibit noticeably smaller variance, and they are closer to the original EoS than any other algorithm, in most cases. Moreover, when 16 M-R points are used instead of 8, they manage to offer a significant improvement in performance. However, cases of moderate results may occur, where the predictions do not exactly follow the straight line of the original EoS. As proof of this statement, we present our reconstruction attempt of the EoSs {\bf MITbag-18} and {\bf MITbag-134} (see Fig.~\ref{MITbag_predict}), as well as the reconstruction attempt of the EoSs {\bf CFL-8} and {\bf CFL-418} (see Fig.~\ref{CFL_predict}). We speculate the drop in performance, might occur due to the overlap area of the \textit{MITbag} and \textit{CFL} curves. When an $M-R$ curve is provided from that area, the algorithms cannot clearly distinguish which type of EoS to predict (\textit{MITbag} or \textit{CFL}) and produce poorer results. 


\section{Concluding Remarks}

The main conclusions of the present study can be summarized as follows

\begin{enumerate}

\item  We applied various machine learning and deep learning methods to reproduce the equation of state of neutron and quark stars, given the mass–radius diagram, also known as the relativistic inverse stellar structure problem. We found that the accurate reproduction of the equation of state can be achieved with varying degrees of precision, especially concerning the region corresponding to high pressure (energy density) values.

\item This study provides the prospect of enabling one, given observational data on the masses and radii of compact objects, to predict or at least obtain a reasonably accurate estimate of the equation of state that characterizes them. Clearly, the greater the amount and precision of the observational information, the more reliable the resulting prediction will be.

\item Another prospect is to be able to directly infer from observational results whether the observed compact object belongs to the category of neutron star quark star or others. This method, which can be characterized as a classification approach, will significantly aid in categorizing the observed compact objects, even if only probabilistically. This could be implemented using appropriate supervised machine learning and Deep learning models. In this way, it enhances our understanding and provides observers with an additional research tool. Therefore, a study in this direction is in order.

\item Undoubtedly, the present study offers ample room for further refinement and modernization. Enhancements may pertain both to the machine learning and neural network methodologies employed and to the quality and breadth of the data, concerning compact objects. In particular, the inclusion of additional observables such as macroscopic parameters, tidal deformability, and microscopic ones such as central density or pressure could significantly improve the accuracy of the results and provide powerful means for the reliable identification of observed compact objects.

\end{enumerate}





\end{document}